\documentclass[aps,prc,preprint,tightenlines,groupedaddress,nofootinbib,
showpacs,preprintnumbers,amsmath,amssymb,superscriptaddress]{revtex4}
\usepackage{graphicx}     % Include figure files
\usepackage{dcolumn}     % Align table columns on decimal point
\usepackage{bm}              % bold math
\usepackage{amsmath}
\begin{document}

\title{Separable Expansions of $V_{low}$ for 2- and 3-Nucleon Systems}
\author{J. R. Shepard}
\affiliation{Department of Physics, University of Colorado, 
	Boulder CO 80309}
\author{J. A. McNeil}
\affiliation{Department of Physics, Colorado School of Mines, 
	Golden CO 80401}
\date{\today} 
\begin{abstract}
We present an alternative organizational scheme for developing effective theories of 2- and 3-body systems that is systematic, accurate, and efficient with controlled errors.  To illustrate our approach we consider the bound state and scattering properties of the 2- and 3-nucleon systems.  Our approach combines the computational benefits of using separable potentials with the improved convergence properties of potentials evolved with a renormalization group procedure.  Long ago Harms showed that any potential can be expanded in a series of separable terms, but this fact is only useful if the expansion can be truncated at low order.  The separable expansion provides an attractive organizational scheme that incorporates the two body bound state in the leading term while allowing for systematic corrections thereafter.  We show that when applied to a renormalization group-evolved potential, the separable expansion converges rapidly, with accurate results for both 2- and 3-body scattering processes using only two separable terms.
\end{abstract}
\maketitle 

\section{Introduction}
\label{Sec:Intro}

Effective Theories (ET's) provide a potent and practical approach to understanding low-energy nuclear few-body systems where it is assumed that the low-energy long distance features should not depend on details of the short distance physics~\cite{Lepage:1990,Georgi:1994,Kaplan:1995uv,Kaplan:1996xu,Kaplan:1998we,vanKolck:1999mw}.  Most contemporary implementations are based on a field theory specified by a Lagrangian containing a series of contact interactions with increasing powers of momentum consistent with the symmetries of the underlying theory. This series, it is hoped, may be truncated at some low order and still account for the low energy physics once the (few) parameters of the ET are fixed by external inputs. The power of the approach is that these few inputs yield a theory which can then be used to calculate systematically any low energy property of the system with relative ease.

However, the use of contact interactions as a starting point, while nearly ubiquitous, is not essential. The original motivation for this choice was presumably one of simplicity. If the physics to be described were perturbative in nature, the organization of the theory would be trivial as terms in the series of contact interactions would appear at the order of their naive scaling dimension. However,  the presence of a weakly bound 2-body state in the $^3S_1$ channel -- the deuteron -- renders the problem fundamentally non-perturbative and the organization of contributions from contact terms in the ET's is, as some practitioners acknowledge, far from simple and often controversial. Also, starting with contact interactions appears to be awkward if one wishes to treat finite range effects in the 2-body interaction. Indeed, Phillips, Beane, and Cohen~\cite{Phillips:1998} have shown that a 2-body potential consisting of contact terms and derivatives thereof necessarily leads to a negative range parameter in the Effective Range Expansion (ERE) of $p \cot(\delta)$ regardless of the number of terms in the expansion.  Such considerations lead one to consider alternative organizational principles that acknowledge the finite range of the 2-body interactions at the outset.  One such approach is based on treating the 2-body interaction as a sum of separable terms, each consisting of the outer product of  {\it form factors} which reflect the structure of real or virtual 2-body states. 

Separable potentials have a long history in nuclear physics going back to the period immediately following Faddeev's seminal work on the 3-body problem when calculations of $Nd$ scattering relied on finite-range separable $NN$ interactions to an extent that was as nearly universal as the use of contact interactions in ET's is today.  More recently, Lepage \cite{Lepage:1997cs} provided a general approach to developing effective interactions starting with the now-canonical series of contact interactions but morphing into a similar series smeared in configuration space by what amounts to a form factor.  The possibility of using separable potentials is also mentioned in passing in several of the important papers in the recent history of ET's addressing three-body physics including Afnan and Phillips \cite{Afnan:2003bs} .   In any case, there is much current interest in how best to treat finite range effects that may enter in Efimov physics which is now accessible experimentally in cold atom systems where one exploits Feshbach resonances to tune the 2-body scattering length, $a$, to infinity. In this regime it would seem prudent to use an approach that does not implicitly assume range effects are negligible in leading order.

To establish the methodology in a simple context we use the uncoupled $^3S_1$ potential of Malfliet and Tjon  (MT-III)~\cite{Malfliet:1968tj} which enables us to treat the 3-nucleon quartet ($^4S_{\frac{3}{2}}$) channel, while avoiding complications due to tensor coupled phase shifts and Efimov physics.  Extension to the doublet case will be treated in a separate paper.  In this work, we exploit the work of Harms \cite{Harms:1970} who showed that a general 2-body potential can be written as an infinite sum of separable terms.  Of course, the utility of the approach depends critically on the convergence properties of the expansion, a study of which is the focus of this work.  There exist fully general Faddeev treatments of the $nd$ system based on realistic microscopic potentials, see ~\cite{Huber:1993, Kievsky:2004-737} and references therein.  The goal of this work is to provide a computationally simpler yet accurate alternative approach exploiting both separable expansions and renormalization group methods.

We consider here three approaches for implementing the Harms expansion.  First, we directly expand the bare 2-body potential to obtain the separable representation of the potential. We find the convergence of the separable expansion for the bare potential is reasonable, but only if a judicious, but somewhat arbitrary, ordering of the separable terms is made.  In addition the Harms form factors for the bare interaction have considerable support at large momenta requiring a large cut-off in the loop integral.  Problems with non-physical negative imaginary phase shifts arise when the loop integration cut-off is reduced.  This suggests that one may profit by first integrating out the high momentum degrees of freedom using a renormalization group scheme, such as the Similarity Renormalization Group (SRG).

The use of the RG in nuclear calculations has been extensively explored in a series of papers by Bogner and collaborators~\cite{Bogner:2005,Bogner:2006a, Bogner:2006pc, Bogner:2007qb}.  More specifically, Bogner, Furnstahl, and Perry ~\cite{Bogner:2006pc} have recently shown how the SRG drives the matrix elements of the bare $NN$ interaction to the diagonal and, in a subsequent paper, used the SRG in the analysis of a 2-state model 3-body problem~\cite{Bogner:2007qb}.   Thus, our second approach is to first evolve the bare $NN$-potential to some low momentum scale before expanding the resulting $V_{low}$ using the methods of Harms.  We find that the convergence properties of the separable expansion are much improved, requiring only two terms to reproduce the exact 2-body T-matrix out to about 400 MeV/c and suffering no issues with negative imaginary phase shifts in 3-body applications.  

Our third approach is to develop a phenomenological separable potential (dipole or gaussian times polynomial form) and fix its parameters by fitting to the 2-body phase shifts.  In an earlier work~\cite{Shepard:2009}, we used this approach by truncating the separable expansion at leading order (an approximation historically referred to as the ``Unitary Pole Approximation" ).  We then used a phenomenological dipole form factor with parameters fit to the phase shifts.  By expanding this model in the inverse of the dipole cut-off parameter, we were able to study systematically finite range expansions, to make connection to the modern ERE's based on contact interactions, and thereby to reveal limitations of the ERE in accounting for finite range effects in the $Nd$ system.   Kamada, et al. also studied 3-nucleon systems using separable expansions of low-momentum effective potentials using a modified Suzuki and Okubo approach to generate $V_{low}$, but using a phenomenological separable expansion in Legendre polynomials~\cite{Kamada:2005hr}. Their expansion converged relatively slowly; requiring as many as 16 terms in the separable expansion. 

We next move to an examination of the 3-nucleon problem, confining our attention to the relatively simple quartet, $^4S_{\frac{3}{2}}$, case to develop our methodology.   We confirm the utility of combining the SRG and the Harm's separable expansion by calculating three nucleon quartet phase shifts,  including the sensitive and small imaginary phase shifts, accurately using only rank-2 approximation.  We emphasize that the approximation method is general, systematic, and controlled.

The paper is organized as follows. In Section II we review Harms' derivation of the separable expansion which defines our notation. We apply the method to the Malfliet-Tjon (MT-III) potential,Ref.~\cite{Malfliet:1968tj}, and show its convergence properties as the rank of the expansion is increased.  In Section III we review the Similarity Renormalization Group and then use it to flow the bare MT-III $NN$ potential to low momentum.  In doing so, as also shown by Bogner et al., we illustrate how the SRG suppresses the off-diagonal components of the potential at high momentum thereby effectively rendering the high momentum components of $V_{low}$ sterile to scattering.  We conclude Section III by expanding the SRG-evolved potential in a separable expansion and show that the convergence properties are improved over that of the bare potential.  In Section IV we turn to the 3-body problem by treating $Nd$ quartet scattering. We first review our UPA results and then examine those of the rank-2 separable approximation to the SRG evolved MT-III potential. The agreement with the full Faddeev calculations of Huber, et al.~\cite{Huber:1993} and Kievsky, et al. \cite{Kievsky:2004-737} is excellent.  We discuss our results and prospective future developments in the concluding section.

\section{Harms' Separable  Expansion}
\label{Sec:Harms}

In the early days of nuclear 3-body calculations, practitioners made use of a simple separable expansion of the 2-body interaction to render the 2-body problem in anaytic form~\cite{Weinberg:1963a,Lovelace:1964,Aaron:1966,Ball:1968,Sitenko:1968,Fuda:1969a,Fuda:1969b}. These applications employed the simplest form of a separable interaction, namely one consisting of a single term which, in terminology later, is a {\it rank-1} separable potential. These early papers  emphasized that such an approximation is well-suited to situations where the 2-body properties are dominated by a single real or virtual bound state, or {\it dimer}.  Use of a rank-1 interaction was typically referred to as the Unitary Pole Approximation (UPA) since, as will be shown below, 2-body unitarity is guaranteed using this form of the interaction.

In his seminal 1963 paper, Weinberg~\cite{Weinberg:1963a} showed how the eigenvalues of the operator $G_0(E) V$ can be used as an organizing principle for developing a perturbation expansion of the T-matrix.  In 1970, using this organizing principle, E.~Harms showed how {\it any} 2-body interaction can be expressed exactly as an infinite sum of separable terms.  Truncation of the expansion at some finite rank allows the construction of the fully off-shell 2-body $T$-matrix by inverting a matrix with a dimension of the rank retained. His expansion yields the UPA when a single term (associated with the bound state) is retained. It should be noted that since separable expansions are always compact, they cannot converge for local potentials which are non-compact~\cite{Osborn:1973a,Osborn:1973b}.  Nevertheless, as pointed out by Koike~\cite{Koike:1990}, a separable expansion may approximate local potential quite well in limited regions of momentum and energy, and specifically for applications to 3-body problems the Green's function in the connected-kernel 3-body equation regulates the integrand rendering the separable expansion convergent for this important case.   We now briefly reconstruct Harms' arguments using his notation with one exception specified below.

Consider the 2-body Schroedinger equation ($\hbar$, $c=1$, $m_{neutron}=m_{proton}=m$):
\begin{equation}
G_0^{-1}(E_n)|\chi_n\rangle =V|\chi_n\rangle
\label{twobdySE}
\end{equation}
where $G_0(E)$ is the free 2-body propagator. Assume there exists a single bound state at $E\rightarrow -B=-\gamma^2/m$. The bound state satisfies
\begin{equation}
G_0^{-1}(-B)|\chi_B\rangle =V|\chi_B\rangle .
\label{twobdybnd1SE}
\end{equation}
Defining the {\it form factors} via the operation of $G_0^{-1}(-B)$:
\begin{equation}
|\psi_n\rangle=G_0^{-1}(-B)|\chi_n\rangle.
\label{defFF}
\end{equation}
We see that the bound state Schroedinger equation can be written as
\begin{equation}
V G_0(-B)|\psi_B\rangle =\lambda_B |\psi_B\rangle 
\label{twobdybnd2SE}
\end{equation}
where $\lambda_B=1$. Solutions $|\psi_n\rangle$ of the generalization of Eq.~\ref{twobdybnd2SE}
\begin{equation}
V G_0(-B)|\psi_n\rangle =\lambda_n |\psi_n\rangle
\label{twoFFeqn}
\end{equation}
form a complete set of kets with normalization condition
\begin{equation}
\langle\psi_n|G_0(-B)|\psi_n\rangle=-\delta_{n,m} .
\label{FFnorm1}
\end{equation}

%%%%%%%%%%%%%%%%%%%%%%%%%%%%%%%%%%%%%%%%%%%%%%
\begin{figure}[ht]
\vspace{0.50in}
\includegraphics[width=6in,angle=0]{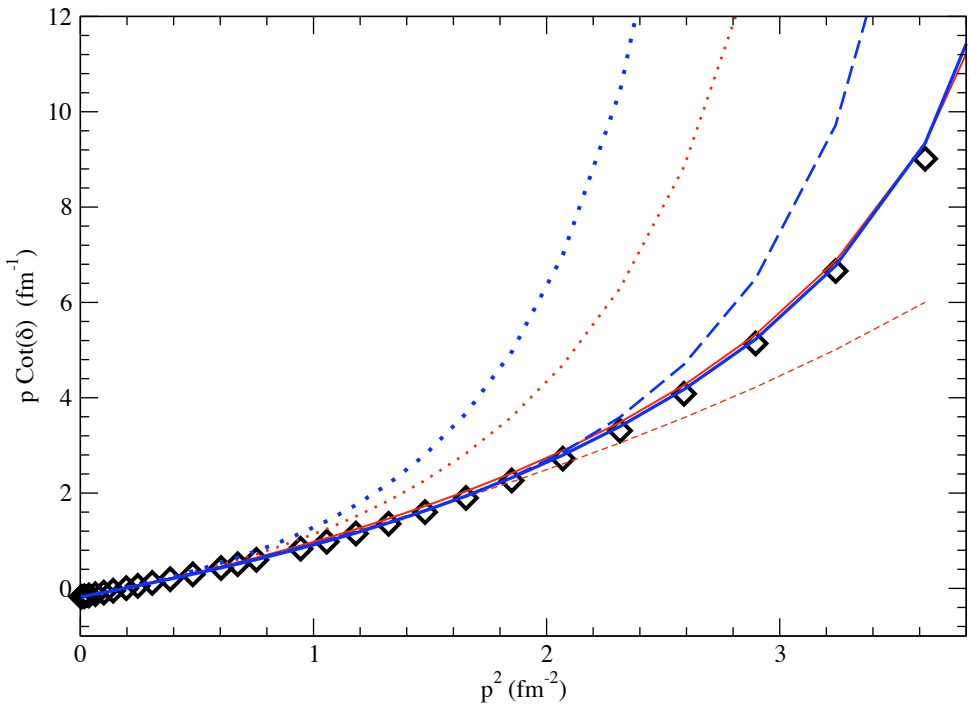}
\caption{(Color online).
         $p \cot(\delta)$ in the $^3S_1$ channel for potential MT-III. The black diamonds are the full Lippmann-Schwinger (LS) calculation taken to be exact. The light dotted (red), short dashed (red), and light (red) lines are the rank-1 (UPA), rank-2, rank-4 results, respectively, for the Harms separable expansion of the bare potential, $V_0$. The heavy dotted (blue), long dashed (blue), and solid (blue) lines lines are the rank-1 (UPA), rank-2, and rank-4 results, respectively, for the Harms separable expansion of the SRG-evolved potential, $V_{low}$.}
\label{Fig1}
\end{figure}
%%%%%%%%%%%%%%%%%%%%%%%%%%%%%%%%%%%%%%%%%%%%%%

We call the $|\psi_n\rangle$'s the ``form factors'' of the potential $V$. Note that our $\lambda_n$'s are the {\it inverses} of those defined by Harms. Also note that, with this normalization convention, the bound state wave function is {\it not} normalized. We may define the wavefunction renormalization $Z_B$ via
\begin{equation}
Z_B^{-1}=\langle\chi_B |\chi_B\rangle= \langle\psi_B|G_0(-B)^2|\psi_B\rangle=-\frac{d}{dE} \bigg|_{E=-B} \langle\psi_B|G_0(E)|\psi_B\rangle .
\label{Zbnd}
\end{equation}
When there is a single bound state, the eigenvalues satisfy
\begin{equation}
\lambda_{n\neq B}<\lambda_B =1.
\label{FFnorm2}
\end{equation}
Weinberg noted that the eigenvalues can be used to develop a perturbative expansion of the T-matrix which will converge when $|\lambda|<1$~\cite{Weinberg:1963a}.  However, for the MT-III potential there are large negative eigenvalues; so the convergence of an expansion based on this ordering principle is in question. We observe that the potential can be written as
\begin{equation}
V=-\sum_n\ |\psi_n\rangle \lambda_n \langle\psi_n|
\label{vFFexp}
\end{equation}
which is verified by substitution into Eq.~\ref{twoFFeqn} and then using the normalization condition~\ref{FFnorm1}. The 2-body $T$-matrix satisfies the Lippmann-Schwinger (LS) equation
\begin{equation}
T({\bf p},{\bf p}';E^+)=V({\bf p},{\bf p}')+\int\frac{d^3q}{(2\pi)^3}\ \frac{V({\bf p},{\bf q})\ T({\bf q},{\bf p}';E)}{E^+ -q^2/m}
\label{LSeq1}
\end{equation}
It is easy to show that
\begin{equation}
T({\bf p},{\bf p}';E^+)=-\sum_{n,m}\  \langle{\bf p} |\psi_n\rangle\ \Delta(E^+)_{n,m}\ \langle\psi_m|{\bf p}'\rangle ,
\label{LSeq2}
\end{equation}
where
\begin{equation}
[\Delta^{-1}(E^+)]_{n,m}=\lambda_n^{-1} \delta_{n,m}+\langle\psi_n|G_0(E^+)|\psi_m\rangle .
\label{DeltaDef}
\end{equation}
Thus, we see that the 2-body T-matrix associated with a separable potential to a given rank involves inverting a matrix with dimension of that rank.  The UPA corresponds to the single term,
\begin{equation}
V_{UPA}\rightarrow -|\psi_B\rangle\  \langle\psi_B|
\label{vUPA}
\end{equation}
where $\lambda_B=1$ has been used, and
\begin{equation}
T_{UPA}({\bf p},{\bf p}';E^+)\rightarrow -\langle{\bf p}|\psi_B\rangle\ \Delta_{UPA}(E^+)\ \langle\psi_B|{\bf p}'\rangle
=-\frac{\langle{\bf p}|\psi_B\rangle\ \langle\psi_B|{\bf p}'\rangle}{1+ \langle\psi_B | G_0(E^+)|\psi_B\rangle}
\label{LSeq3}
\end{equation}
which we recognize as a standard result for (rank-1) separable 2-body interactions \cite{Watson:1967}. The residue of the T-matrix at the deuteron pole, $p\rightarrow\imath\gamma$, is related to the wavefunction normalization, Eq.~\ref{Zbnd}.
\begin{equation}
{\cal R}[T_{UPA}] |_{p\rightarrow \imath\gamma}\rightarrow |\langle{\bf \imath\gamma}|\psi_B\rangle|^2 Z_B.
\label{Residue}
\end{equation}

%%%%%%%%%%%%%%%%%%%%%%%%%%%%%%%%%%%%%%%%%%%%%%
\begin{table}
\begin{tabular}{|l|c|c|}
 \hline
Potential & $\lambda>0$ &$\lambda<0$\\
\hline
$V_0$ &    1.0000&\\
            &    0.159190&  -1.8917\\
             &   0.061669&  -0.24285\\
             &   0.032467& -0.018806\\
\hline
$V_{low}$& 1.0000&\\
                 & 0.11937 & -0.11781\\
                 & 0.024633& -0.05982\\
                 & 0.0055180& -0.050676\\
\hline
\end{tabular}
\caption{First few positive and negative Harms eigenvalues for $V_0$ and $V_{low}$ for the MT-III potential. }
\label{Table1}
\end{table}
%%%%%%%%%%%%%%%%%%%%%%%%%%%%%%%%%%%%%%%%%%%%%%

%%%%%%%%%%%%%%%%%%%%%%%%%%%%%%%%%%%%%%%%%%%%%%
\begin{figure}[ht]
\vspace{0.50in}
\includegraphics[width=6in,angle=0]{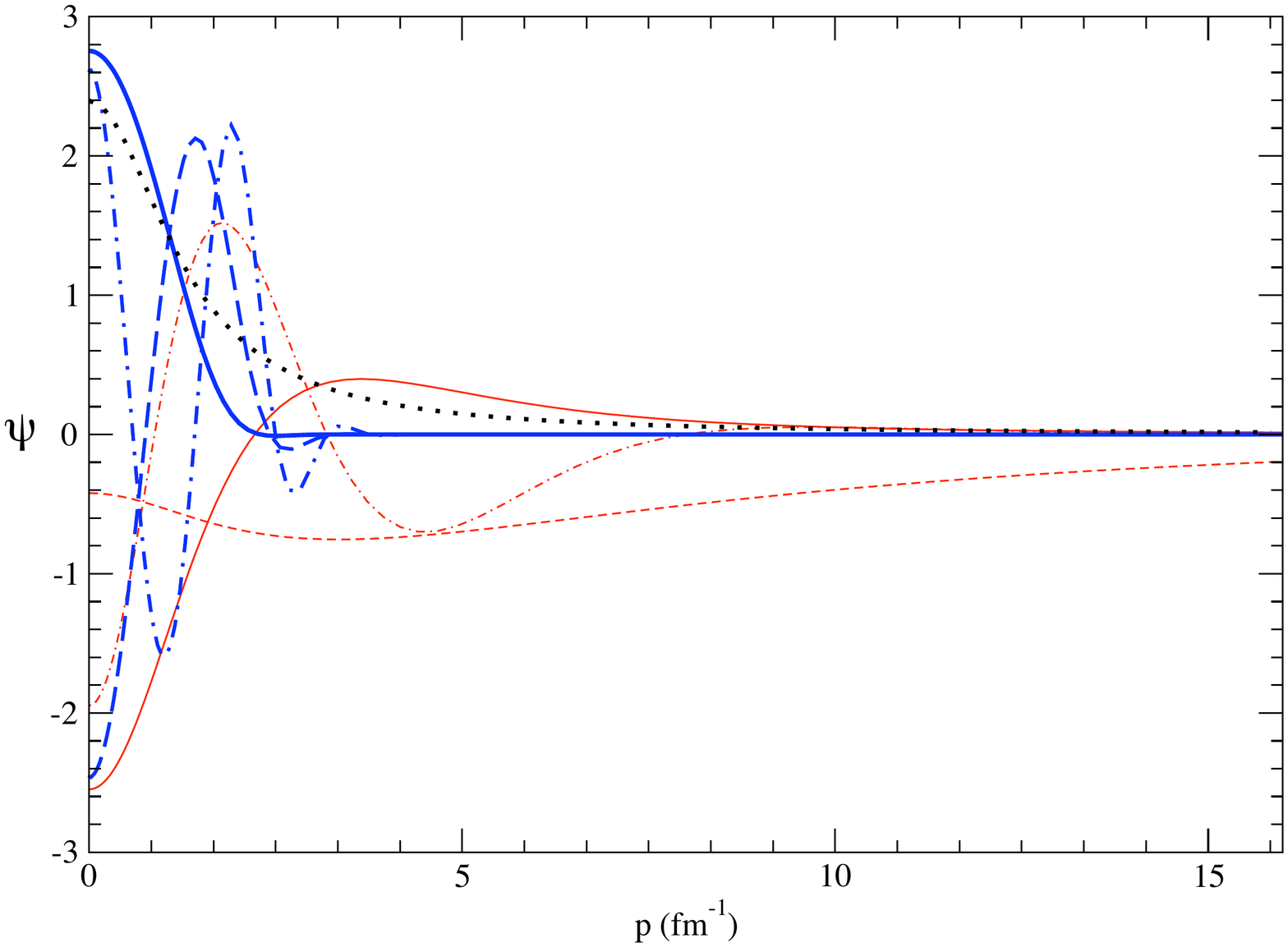}
\caption{(Color online).
         Form factors, $\langle p |\psi_n\rangle$, from the solution of Eq.~\ref{twoFFeqn} for the $^3S_1$ channel. The light solid (red) curve is the bound state form factor ($\lambda=1.00000$) and the short dash (red) and short dash-dot (red) curves are the second ($\lambda=0.159190\dots$) and third ($\lambda=-1.891665\dots$) form factors for the Harms separable expansion of the bare potential, $V_0$.  The thick solid (blue) curve is the bound state form factor ($\lambda=1.00000$), the long dash (blue) curve is the second ($\lambda=0.1193678\dots$) form factor, and  the long dash-dot (blue) curve is the third ($\lambda=0.024633\dots$) form factor for the Harms separable expansion of the SRG potential, $V_{low}$.  Note the strong supression of the SRG- (blue) evolved form factors for $p>\Lambda_{low}=2$ fm$^{-1}$.  For comparison we also show the dotted (black) curve is the phenomenological dipole form factor.}
\label{Fig2}
\end{figure}
%%%%%%%%%%%%%%%%%%%%%%%%%%%%%%%%%%%%%%%%%%%%%%

As with our predecessors, our interest in separable forms of 2-body interactions stems from the great simplification they afford in solving 3-body Faddeev equations (\cite{Watson:1967}).  In the context of the Harms expansion, when focussing on low energy properties of 3-body systems, the promised simplicity is only realized if the separable expressions for $V$ and $T$ given above can be truncated at some low order and still be accurate over the relevant range of momentum and energy. We now examine this issue using the familiar $^3S_1$ $NN$ potential of Malfliet and Tjon~\cite{Malfliet:1968tj}. We use this potential because we consider the present work to be a ``proof of principle'' exercise, and we wish to avoid the complexities of the full coupled channels nature of the $^3S_1 - ^3D_1$ system while treating the quartet channel in the $nd$ system where only the $NN$ triplet channel contributes and complexities involving Efimov physics are absent.

In Figure~\ref{Fig1}, we compare $p\cot\delta$ for Potential III of Malfliet and Tjon (MT-III) computed by exact solution of the LS equation (Eq.~\ref{LSeq1}) with results of the $T$-matrix (Eq.~\ref{LSeq2}) using the Harms method for various truncations.  Ignore for now the curves involving ``$V_{low}$".  We always take the leading term to be that associated with the bound state, $\lambda_1\rightarrow\lambda_B=1$.  We order the separable expansion by the eigenvalues, $\lambda_n$, of the operator $G_0V$, Eq.~\ref{twoFFeqn}, separately tracking positive and negative terms, see Table~\ref{Table1}.  Weinberg showed this forms the basis for a systematic perturbation expansion in that it orders the separable terms according to the magnitude of their contribution to the potential~\cite{Weinberg:1963a,Harms:1970}.  Figure 1 shows the convergence of the separable expansion of the bare potential as the rank is increased, first from rank-1 (UPA), rank-2, fully converging by rank-4. (By a judicious, but arbitrary, choice of ordering, convergence can actually be achieved at rank-3.)  We see that for either potential the rank-4 Harms results are essentially in perfect agreement with the full LS solution over the momentum range shown ($p^2\leq 4\ {\rm fm}^{-2}$).  (This finding is of some interest in its own right as the numerical cost of the Harms calculation is significantly less than that of the full LS computation.)  

Figure 2 shows the form factors for the first few terms in the Harms expansion of the bare potential.  As can be seen, aside from the first (UPA) term, all the higher rank form factors have considerable support at large momenta.  (Ignore for now the curves involving ``$V_{low}$".)  Truncating at rank-1 means that the calculated phase shift will go to zero too rapidly with $p^2$ giving a $p\cot \delta$ consistently above the exact value as shown in Figure 1. To capture the high momentum dependence therefore requires that higher order terms be retained in the Harms expansion.  As we are interested only in the low momentum physical properties, this suggests that a renormalization group procedure could improve the convergence properties of the Harms expansion.  This possibility is explored in the next section.

\section{Determination of $V_{low}$ {\it via} the Similarity Renormalization Group}
\label{Sec:SRG}

Much recent work has focussed on using a renormalization group approach to develop effective 2-body interactions -- designated as $V_{low}$ -- evolved from ``bare'' interactions defined at an energy scale much higher than the scale of the physical application in which the 2-body interaction is to be used.  Several methods for obtaining $V_{low}$ from the bare potentials have been developed over the years including the Kuo-Kurasawa (KK), Lee-Suzuki (LSuz), Contractor Renormalization (CORE) and Similarity Renormalization Group (SRG) approaches.  We have performed calculations using each of the last three methods and find similar results but here focus on the SRG whose basic features we now review.

The basic concepts of the SRG are straight-forward; they were formulated independently in the early 1990's by Wegner~\cite{Wegner:1994} and by Glazek and Wilson~\cite{Glazek:1993,Glazek:1994}. One begins with a Hamiltonian for the 2-body interaction, $H_0=T+V_0$. For simplicity, we take a basis of momentum eigenstates with a maximum momentum, $\Lambda_0$, assumed to be very large compared to the momentum scale of the 2-body physics of interest. Finding the eigenvalues and eigenvectors of $H_0$ constitutes a full solution of the 2-body problem. The SRG allows determination of $H_{low}=T+V_{low}$ via a sequence of similarity transformations which (i) retains all original eigenvalues while (ii) reducing the upper limit of momentum for which $V_{low}$ has non-trivial dynamical content -- which is identified with  $\Lambda_0$ for $V_0$ -- to some lower momentum scale, $\Lambda_{low}$. Consider the similarity transformation,
\begin{equation}
H(s)=U(s)\ H_0\ U(s)^\dagger,
\label{SRG1}
\end{equation}
where  $s$ is a scale parameter related to momenta $\Lambda$ between $\Lambda_0$ and $\Lambda_{low}$.  Differentiation of Eq.~\ref{SRG1} with respect $s$ yields
\begin{equation}
\frac{d H(s)}{d s}=[\eta(s),H(s)]
\label{SRG2}
\end{equation}
where
\begin{equation}
\eta (s)=\frac{d U(s)}{d s}\ U^\dagger(s)=-\eta(s)^\dagger.
\label{eta1}
\end{equation}
We recognize Eq.~\ref{SRG2} as a Renormalization Group (RG) flow equation for $H(s)$. The matrix $\eta(s)$ specifies the flow and a good choice for $\eta(s)$ in our momentum basis where the kinetic energy operator $T$ is diagonal has been shown to be
\begin{equation}
\eta(s)=[T,H(s)]
\label{eta2}
\end{equation}
which yields the familiar double-commutator form of the SRG flow equation
\begin{equation}
\frac{d H(s)}{d s}=\bigl[[T,H(s)],H(s)\bigr].
\label{SRG3}
\end{equation}
Writing out this equation explicitly in terms of matrix elements of $T$ and $H(s)$, we find
\begin{equation}
\frac{d H_{ij}}{d s}=-(T_{ii}-T_{jj})^2 H_{ij}+\sum_k [T_{ii}+T_{jj}-2 T_{kk}] H_{ik} H_{kj} 
\label{SRG4}
\end{equation}
where we have dropped the explicit dependence of $H$ on $s$. Focussing on the first term on the rhs of this equation, we see that the SRG ``flow'' drives off-diagonal matrix elements of $H$ toward zero at a rate which is quadratic in the distance from the diagonal. This behavior dominates even when the contributions of the second term are included. Thus the SRG flow (i) keeps all original eigenvalues ({\it i.e.}, at $\Lambda_0$) unchanged since successive $H$'s are related by a similarity transformation and (ii)  drives the $H(p,p')$ matrix toward a diagonal form for $p$ or $p'>\Lambda_{low}$. Hence, these high-momentum modes have no {\it dynamical properties}; {\it e.g.}, they do not induce scattering! The remaining dynamics for modes below $\Lambda_{low}$ embody the effects  of the high-momentum modes indirectly in the general spirit of the Renormalization Group method.

%%%%%%%%%%%%%%%%%%%%%%%%%%%%%%%%%%%%%%%%%%%%%%
\begin{figure}[ht]
\vspace{0.50in}
\includegraphics[width=6in,angle=0]{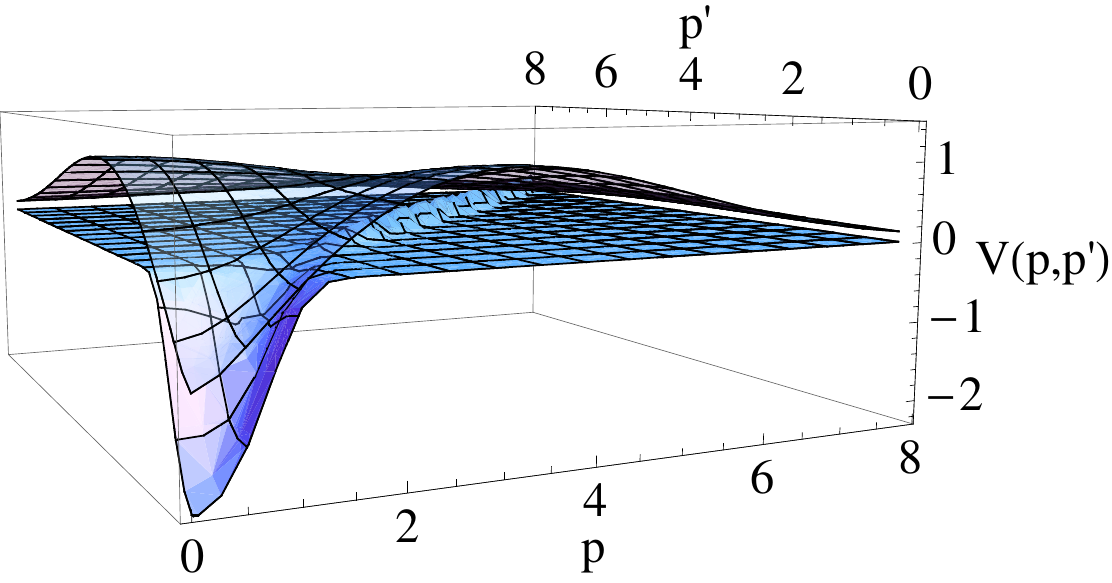}
\caption{(Color online).
         The MT-III potential ($\times\ m/\hbar^2$) in fm before ($V_0(p,p')$, upper contour) and after ($V_{low}(p,p')$, lower contour) SRG evolution from $\Lambda_0=16\ {\rm fm}^{-1}$ to $\Lambda_{low}=2\ {\rm fm}^{-1}$.}
\label{Fig3}
\end{figure}
%%%%%%%%%%%%%%%%%%%%%%%%%%%%%%%%%%%%%%%%%%%%%%

It is evident that $s$ has units of $m^{-4}$; so we set $s_0=\Lambda_0^{-4}$ and $s_{low}=\Lambda_{low}^{-4}$. In what follows, we take the MT-III potential for $V_0$ and $s_0$ and $s_{low}$ were chosen to coincide with $\Lambda_0=16\ {\rm fm}^{-1}$ and $\Lambda_{low}=2\ {\rm fm}^{-1}$, respectively. We obtain $V_{low}$ by numerically integrating Eq.~\ref{SRG4} from $s_0$ to $s_{low}$ using a 4th order Runge-Kutta method or by using the Cayley approximation for the infinitesimal similarity transformation. Both methods give equivalent results.  3D plots of $V_0$ and $V_{low}$ are displayed in Fig.~\ref{Fig3} where $V_{low}$ has substantial off-diagonal contributions only below $\Lambda_{low}$. For larger momentum the off-diagonal components are suppressed and there is significant strength only along the diagonal which renders the potential sterile to scattering at larger momenta.  These results confirm the general behavior reported by Bogner, Furnstahl, and Perry~\cite{Bogner:2007qb}.  At low momentum $V_{low}$ is substantially deeper than $V_0$. 

Next we perform a separable expansion of $V_{low}$.  The differences between the separable expansion of $V_0$ and $V_{low}$  are readily apparent in the Harms form factors of Fig.~\ref{Fig2} where we show the first two form factors arising from the solution of Eq.~\ref{twoFFeqn} with $V\rightarrow V_{low}$. Fig.~\ref{Fig2} also shows the same quantities for $V\rightarrow V_0$, and reveals that the SRG suppresses the form factors of $V_{low}$ at momenta larger than $\Lambda_{low}$.  Moreover, there are no longer any form factors with large negative eigenvalues. The form factor associated with the bound state always has eigenvalue $\lambda_1=1$, of course, while the second has eigenvalue for $V_{low}$, $\lambda_2 = 0.119368\dots$, is considerably different from the large negative value of $\lambda_2 = -1.89166\dots$ obtained with $V_0$.  Also, with $V_{low}$, all remaining eigenvalues have magnitudes less than $\lambda_2$, e.g. $\lambda_3=0.024633\dots$, and so on.  (The SRG form factor associated with the negative eigenvalue, $\lambda=-0.11781\dots$, vanishes for $q<\Lambda_0$ and is ignored.) This forms the basis for a systematic organization of terms in what can be considered a rapidly converging perturbation expansion~\cite{Weinberg:1963a,Bogner:2005}. 

The calculations of the 2-body $p \cot\delta$ resulting from the rank-1 (UPA), rank-2, and rank-4 expansions of $V_{low}$ appear in Fig.~\ref{Fig1} along with the same results for $V_0$. We see that the separable expansion of $V_{low}$ for momenta below $\Lambda_{low}=2$ fm$^{-1}$ is essentially converged by rank-2 which is a modest improvement over the equivalent expansion for $V_0$ while, by the nature of the similarity transformation, leaving the deuteron binding energy unchanged. These findings encourage us to employ the equivalent of the rank-2 $T$-matrix for $V_{low}$ in the 3-body, $nd$, elastic scattering calculations considered in the next section.  As that discussion will indicate, for 3-body applications, we must be able to analytically continue the 2-body $T$-matrix away from the real axis without introducing spurious poles. We accomplish this by fitting the relevant form factors shown in Fig.~\ref{Fig2} with sums of even polynomials of $p$ times Gaussians. The fits we have obtained are excellent and would be indistinguishable from the numerical results in Fig.~\ref{Fig2}.

\section{Unitary Pole Approximation with a Dipole Form Factor}
\label{Sec:UPA}

Before developing the higher rank separable expressions and to provide a model with which to compare, we briefly review the example of a rank-1 separable 2-body $s$-wave treatment using a phenomenological dipole form factor closely following our earlier work~\cite{Shepard:2009}. We limit our development to s-waves, but the forms easily generalize to the higher partial waves.  Recalling Eq.~\ref{vUPA}, 
\begin{equation}
V\rightarrow -|\psi_B\rangle\    \langle\psi_B|
\label{Twobodypot}
\end{equation}
where $\lambda_B=1$ has been used.  Now using a Yamaguchi or dipole form factor~\cite{Yamaguchi:1954}, we have 
\begin{equation}
\langle {\bf p}|\psi_B\rangle={\cal N}^{1/2} g(p) =\frac{{\cal N}^{1/2}}{(1+p^2/\beta^2)}
\label{Dipole1}
\end{equation}
where $g(p)$ is the form factor which satisfies $g(0)=1$,  $p=|{\bf p}|$, and $\cal N$ is found by normalizing according to Eq.~\ref{FFnorm1},
\begin{equation}
{\cal N}^{-1}=-\langle\psi_B|G_0(-B)|\psi_B\rangle=\int \frac{d^3p}{(2\pi)^3} \frac{g(p)^2}{B+q^2/m}=\frac{m}{8\pi}\frac{\beta^3}{(\beta+\gamma)^2}
\label{DipoleNorm}
\end{equation}
where $\gamma=\sqrt{m B}$ has been used.   As discussed in Ref.~\cite{Shepard:2009}, the dipole form factor is intimately related to the Hulthen wavefunction~\cite{Hulthen:1957}.  Similarly, we find for the deuteron propagator,
\begin{equation}
\Delta^{-1}(E^+)= 1-\frac{(\beta+\gamma)^2}{\biggl(\beta+\sqrt{-m E^+}\biggr)^2}.
\label{DeltaUPA}
\end{equation}
 Finally, using the dipole form factor in Eq.~\ref{Zbnd} gives the wave function normalization factor,
\begin{equation}
Z_B^{-1}=\frac{m}{\gamma(\beta+\gamma)} .
\label{ZUPA}
\end{equation}

Using Eqs.~\ref{LSeq3}, ~\ref{Dipole1}, and ~\ref{DeltaUPA}, we construct the fully off-shell elastic 2-body amplitude, $f_2$, via
\begin{eqnarray}
f_2(p,p';E)&=&-\frac{m}{4\pi}\ T(p,p';E)\rightarrow -\frac{m}{4\pi}\biggl(-{\cal N} g(p^2)\ g({p'}^2)\Delta(E)\biggr)\nonumber \\
&=&2\frac{(\beta+\gamma)^2}{\beta^3}\frac{1}{(1+p^2/\beta^2)(1+{p'}^2/\beta^2)}\biggr(1-\frac{(\beta+\gamma)^2}{\biggl(\beta+\sqrt{-m E^+}\biggr)^2}\biggl)^{-1},
\label{Fullfofp}
\end{eqnarray}
and the on-shell scattering amplitude is given by
\begin{equation}
f(p)=f_2(p,p;E=p^2/m+i\eta)=\biggl[\frac{(-\beta^2\gamma(2\beta+\gamma)+(3\beta^2+2\beta\gamma+\gamma^2)p^2+p^4)}{2\beta(\beta+\gamma)^2}-i p\biggr]^{-1}.
\label{fofp}
\end{equation} 
Note that the 2-body amplitude has a pole at $E=-\gamma^2/m$  as it must. We will see later that this pole structure plays a critical role in 3-body calculations by tracking the breakup threshold. In standard fashion we introduce the phase shift, $\delta$, via
\begin{equation}
f(p)=\frac{1}{p\cot\delta -\imath p},
\label{pcotdelta}
\end{equation}
and see that Eq.~\ref{fofp} contains the correct unitarity term and that we can identify
\begin{equation}
%p \cot\delta=\frac{(p^2+\beta^2)^2}{2\beta(\beta+\gamma)^2}+\frac{p^2-\beta^2}{2\beta}.
p \cot\delta=-\frac{\beta\gamma(2\beta+\gamma)}{2(\beta+\gamma)^2}+\frac{3\beta^2+2\beta\gamma+\gamma^2}{2\beta(\beta+\gamma)^2} p^2+\frac{1}{2\beta(\beta+\gamma)^2} p^4.
\label{pcotdel}
\end{equation}
Comparison with the Effective Range Expansion,
\begin{equation}
p \cot\delta=-\frac{1}{a}+\frac{1}{2} r_o p^2+v_4 p^4\dots,
\label{EREdef}
\end{equation}
allows us to identify
\begin{equation}
a=\frac{2(\beta+\gamma)^2}{\beta\gamma(2\beta+\gamma)},\quad r_0=\frac{3\beta^2+2\beta\gamma+\gamma^2}{\beta(\beta+\gamma)^2}\quad {\rm and}\quad v_4=\frac{1}{2\beta(\beta+\gamma)^2} .
\label{EREconsts}
\end{equation}
Note that the ERE terminates at ${\cal O}(p^4)$ for the present choice of $g(p)$.  The exact deuteron binding energy for the MT-III potential is $B=2.23995\dots$ MeV which implies $\gamma=0.232405\dots$ fm$^{-1}$. The numerical solution of the exact Lippmann-Schwinger Equation for this potential yields phase shifts which in turns can be used to compute $p \cot\delta$, a fit to which gives the values:
\begin{equation}
%a^{\rm fit}=5.495\ {\rm fm},\quad r_0^{\rm fit}=1.904\ {\rm fm}\quad {\rm and}\quad v_4^{\rm fit}=0.1028\ {\rm fm}^3 .
a^{\rm fit}=5.505\ {\rm fm},\quad r_0^{\rm fit}=1.891\ {\rm fm}\quad {\rm and}\quad v_4^{\rm fit}=0.110\ {\rm fm}^3 .
\label{EREexact}
\end{equation}
Fixing $\gamma$ to the value given above and determining $\beta=1.281$ fm$^{-1}$ by fitting to $a^{\rm fit}$ then Eq.~\ref{EREconsts} {\it predicts}
\begin{equation}
%r_0=1.885\ {\rm fm}\quad {\rm and}\quad v_4=0.1662\ {\rm fm}^3 
r_0=1.897\ {\rm fm}\quad {\rm and}\quad v_4=0.170\ {\rm fm}^3 
\label{EREfit}
\end{equation}
which is in exceptional agreement with the fitted values considering the simplicity of the approximation. Similar levels of agreement are found using Gaussian form factors, $g(p)=\exp(-p^2/\beta^2)$, but the finite range expansion does not terminate. Note that the value of $\beta$ is roughly two pion masses which we would expect qualitatively for this ``cutoff" parameter; {\it i.e.}, $\beta$ is ``natural". Finally, we observe that none of the quantities in Eq.~\ref{EREconsts} is ill-behaved in the short-range limit ($\beta\rightarrow\infty$), namely, $a\rightarrow1/\gamma$, $r_0\sim\beta^{-1}\rightarrow 0$, and $v_4\sim \beta^{-3}\rightarrow 0$ as expected.

%%%%%%%%%%%%%%%%%%%%%%%%%%%%%%%%%%%%%%%%%%%%%%
\begin{figure}[ht]
\vspace{0.50in}
\includegraphics[width=4in,angle=0]{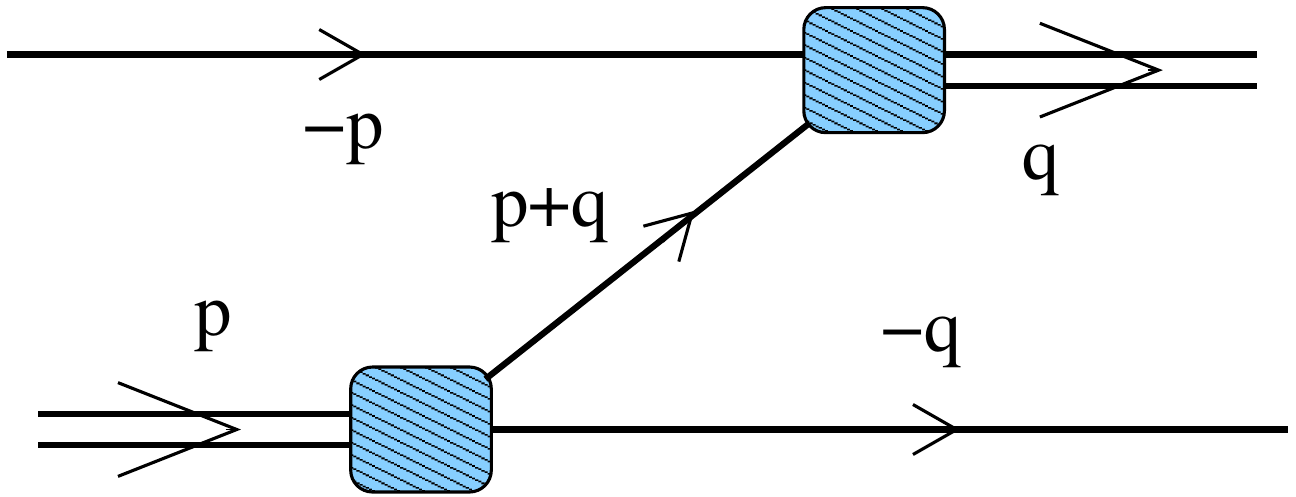}
\caption{(Color online).
         Diagrammatic representation of $\cal Z$, the Born (``pinball") term of 3-body particle-dimer elastic scattering Faddeev equation. The `` blobs'' are the form factors which depend on the magnitudes of the relative 2-body momenta which are $|{\bf p}/2+{\bf q}|$ and $|{\bf p}+{\bf q}/2|$ for the lower and upper form factors, respectively.}
\label{ZDiag}
\end{figure}
%%%%%%%%%%%%%%%%%%%%%%%%%%%%%%%%%%%%%%%%%%%%%%

%%%%%%%%%%%%%%%%%%%%%%%%%%%%%%%%%%%%%%%%%%%%%%
\begin{figure}[ht]
\vspace{0.50in}
\includegraphics[width=6in,angle=0]{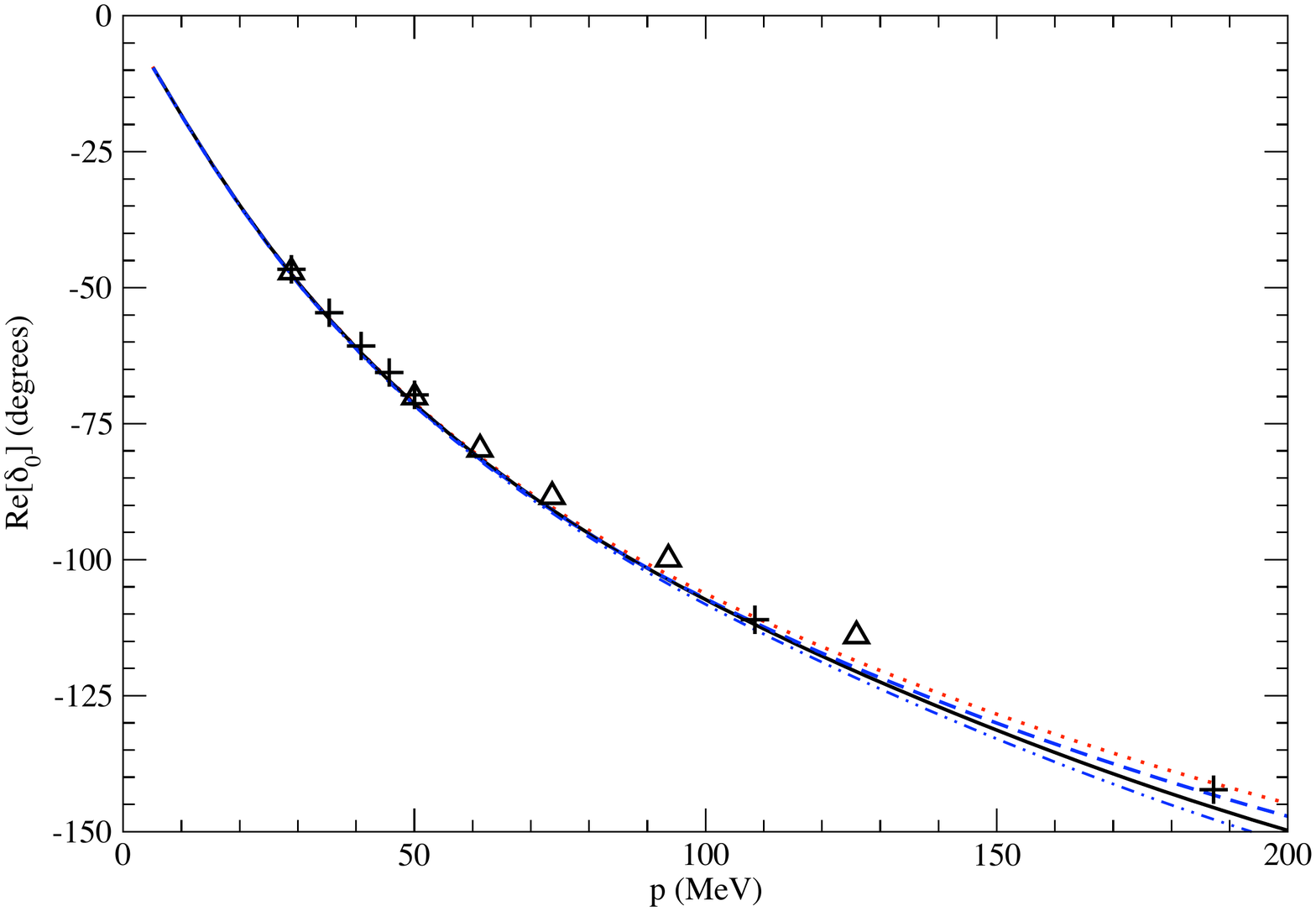}
\caption{(Color online).
The real part of $L=0$ $nd$ quartet phase shifts using three implementations of the rank-1 (UPA) model. The solid curve (black) is the rank-1 calculation for $V_{low}$ and the dashed curve (blue) is the rank-1 calculation for $V_{0}$. The dotted curve (red) is the rank-1 calculation using the phenomenological dipole form factors discussed in Sec.~\ref{Sec:UPA}.  Also shown is the effect of reducing the cut-off reduced from $\Lambda = 16$ fm$^{-1}$ to $\Lambda = 2$ fm$^{-1}$.  This change has no significant effect on the rank-1 calculation for $V_{0}$ and yields the dash-dot-dot curve (blue) is the rank-1 calculation for $V_{low}$.   The triangles are the full Faddeev calculations of Huber, et al.~\cite{Huber:1993} and Kievsky, et al.~\cite{Kievsky:2004-737} and references therein.}
\label{UPA_S_Re}
\end{figure}
%%%%%%%%%%%%%%%%%%%%%%%%%%%%%%%%%%%%%%%%%%%%%%

%%%%%%%%%%%%%%%%%%%%%%%%%%%%%%%%%%%%%%%%%%%%%%
\begin{figure}[ht]
\vspace{0.50in}
\includegraphics[width=6in,angle=0]{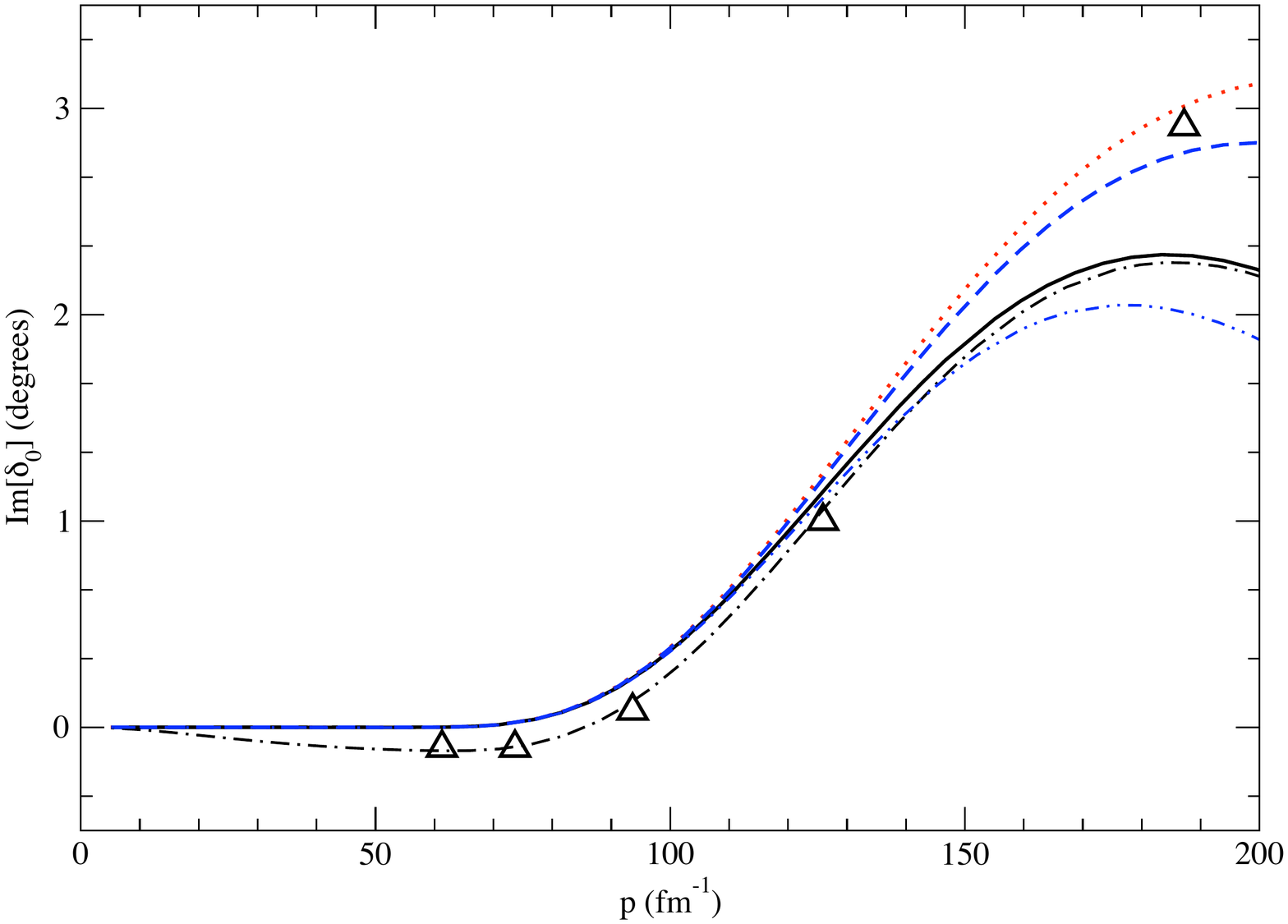}
\caption{(Color online).
         The imaginary part of $L=0$ $nd$ quartet phase shifts using three implementations of the rank-1 (UPA) model. The solid curve (black) is the rank-1 calculation for $V_{low}$ and the dashed curve (blue) is the rank-1 calculation for $V_{0}$. The dotted curve (red) is the rank-1 calculation using the phenomenological dipole form factors discussed in Sec.~\ref{Sec:UPA}.  Also shown is the effect of reducing the cut-off reduced from $\Lambda = 16$ fm$^{-1}$ to $\Lambda = 2$ fm$^{-1}$.  This change gives the dash-dot curve (black) for the rank-1 calculation for $V_{0}$ and the dash-dot-dot curve (blue) for the rank-1 calculation for $V_{0}$.   The triangles are the full Faddeev calculations of Huber, et al.~\cite{Huber:1993}. The breakup threshold is at $p=53$ MeV/c.}
\label{UPA_S_Im}
\end{figure}
%%%%%%%%%%%%%%%%%%%%%%%%%%%%%%%%%%%%%%%%%%%%%%

\section{Faddeev equations for $^4S_{\frac{3}{2}}\ nd$ Scattering}
\label{Sec:ThreeBodyNew}

Having established our method for treating the 2-body interaction, we now address the problem of $nd$ $^4S_{\frac{3}{2}}$ (or ``quartet'') scattering. In an earlier work~\cite{Shepard:2009}, we made use of the Unitary Pole Approximation, rank-1 separable approximation, to study convergence properties and limitations in treating finite range effects using effective range expansions (ERE's). In that work we found that ERE's can account for finite range effects adequately to Next-to-Leading (NLO) order, but fundamentally different terms, similar to 3-body contact terms, arise in the finite range expansion of the UPA at Next-to-Next-Leading order (N$^2$LO) that do not appear in the ERE formulations.  Furthermore, in treating 3-body systems, there are catastrophic failures of unitarity in the form of negative imaginary phase shifts in low order ERE treatments.   Including vertex corrections from explicit pions treated in chiral perturbation theory have been shown to cure this disease~\cite{Bedaque:1999vb}, but at a consider cost in effort and transparency.  

We now construct the 3-body Faddeev equation in the Unitary Pole Approximation.  Consider the scattering of a low energy neutron from a deuteron in the quartet, $^4S_{\frac{3}{2}}$, channel.  Treating the nucleons as identical particles in an isospin formalism, the S-wave ($L=0$) 3-body t-matrix is given by~\cite{Watson:1967}
\begin{equation}
X^{UPA}(p, p';E)= 2 {\cal Z}^{UPA}(p, p';E)-\frac{1}{2 \pi^2} \int~q^2 dq\ 2 {\cal Z}^{UPA}(p, q;E)~\Delta(E-\frac{3 q^2}{4 m})~X^{UPA}(q, p';E),
\label{X_UPA}
\end{equation}
where $\Delta(E)$ is given by Eq.~\ref{DeltaUPA}, identified with the deuteron (or {\it dimer}) propagator, and $\cal Z$ is the particle-exchange (``pinball") amplitude depicted diagrammatically in Fig.~\ref{ZDiag},
\begin{equation}
{\cal Z}^{UPA}(p, p';E)=-\Lambda_{IS} \frac{m}{2} \int_{-1}^{1}dx \frac{\psi_B(|{\bf p +p'/2}|)\psi_B(|{\bf p' +p/2}|)}{p^2+p'^2-m E^+ +p p' x}P_0(x)
\label{Z}
\end{equation}
where $P_\ell(x)$ is a Legendre polynomial and $\Lambda_{IS}$ is the isospin-spin structure factor which is unity for bosonic scattering.  For the $nd$ quartet case,  
\begin{equation}
\Lambda_{IS}=U\biggl(\frac{1}{2},\frac{1}{2},\frac{3}{2},\frac{1}{2};1,1\biggr)\times U\biggl(\frac{1}{2},\frac{1}{2},\frac{1}{2},\frac{1}{2};0,0\biggr)=\biggl(1\biggr)\times\biggl(-\frac{1}{2}\biggr)
\label{SI}
\end{equation}
where the $U$'s are the unitary Racah recoupling coefficients.  Examining the homogeneous terms, we note that by associating the $\psi_B(p)$ factors from ${\cal Z}(p, q;E)$ with the dimer propagator, $\Delta(E-\frac{3 q^2}{4 m})$, the (off-shell) 2-body T-matrix, Eq.~\ref{LSeq3}, is reconstituted.

Including the wavefunction renormalization factor, Eq.~\ref{ZUPA}, defines the off-shell 3-body T-matrix,
\begin{equation}
T_{3}^{UPA}(p,p';E)= Z_B X^{UPA}(p,p';E),
\label{T3Def}
\end{equation}  

The Faddeev equation for $T_3$ becomes,
\begin{equation}
T_{3}^{UPA}(p,p';E)=T_{3}^{Born}(p,p';E)+\int_0^\infty dq q^2 K_3^{UPA}(p,q;E) T_{3}^{UPA}(q,p';E),
\label{T3Eqn0}
\end{equation}
where
\begin{eqnarray}
T_3^{Born}(p,p';E)&=&2 Z_B {\cal Z}( p, p';E),~~{\rm and}\\
K_3^{UPA}(p,q,E)&=& -2\frac{4\pi}{(2\pi)^3}{\cal Z}(p,q;E)~\Delta_{UPA}\biggl(E-\frac{3 q^2}{4 m}\biggr)\nonumber\\
&=& \frac{\Lambda_{IS}} {2\pi^2} \int_{-1}^1 dx ~\psi_B\biggl(\sqrt{p^2+\frac{q^2}{4}+p q x}\biggr)\psi_B\biggl(\sqrt{\frac{p^2}{4}+q^2+p q x}\biggr)~\Delta_{UPA}\biggl(E-\frac{3 q^2}{4 m}\biggr).\nonumber\\
\label{KernelDef}
\end{eqnarray}
Using the dipole form factor, Eq.~\ref{Dipole1}, we can perform the needed angle integration in Eq.~\ref{Z} to find the UPA result,
\begin{eqnarray}
T_3^{Born}(p,q;E)
&=& \frac{32\pi\beta^4\gamma(1+\gamma/\beta)^3\Lambda_{IS}}{3 m p q (p^2-q^2)(3 q^2/4- \beta^2- m E)(3 p^2/4- \beta^2- m E)}\nonumber\\
&\times&\Biggl\{  \frac{3}{4}(p^2-q^2)\ln\biggl|\frac{p^2+q^2-m E+p q}{p^2+q^2-m E-p q}\biggr| \nonumber\\
&-&  (3 p^2/4- \beta^2- m E)\ln\biggl|\frac{4p^2+q^2+4\beta^2+4p q}{4p^2+q^2+4\beta^2-4p q}\biggr| \nonumber \\
&+& (3 q^2/4- \beta^2- m E)\ln\biggl|\frac{p^2+4q^2+4\beta^2+4p q}{p^2+4q^2+4\beta^2-4p q}\biggr|  \Biggr\}.
\label{BornUPA}
\end{eqnarray}
Likewise for the 3-body kernel, we find
\begin{eqnarray}
K_3^{UPA}(p,q,E)&=&   \frac{4 \beta(\gamma+\beta)^2\Lambda_{IS}}{3 \pi m p q} \frac{1}{1-\frac{(\beta+\gamma)^2}{(\beta+\sqrt{3q^2/4-m E})^2}}\nonumber\\
&&\Biggl\{   \frac{3}{(3 p^2/4- \beta^2- m E)(3 q^2/4- \beta^2- m E)} \ln\biggl|\frac{p^2+q^2-m E+p q}{p^2+q^2-m E-p q}\biggr|\nonumber\\
&&-\frac{1}{(p^2-q^2)(3 q^2/4- \beta^2- m E)} \ln\biggl|\frac{4p^2+q^2+4\beta^2+4p q}{4p^2+q^2+4\beta^2-4p q}\biggr| \nonumber \\
&&+\frac{1}{(p^2-q^2)(3 p^2/4- \beta^2- m E)}\ln\biggl|\frac{p^2+4q^2+4\beta^2+4p q}{p^2+4q^2+4\beta^2-4p q}\biggr|  \Biggr\}.
\label{K3UPA}
\end{eqnarray}

The on-shell 3-body S-wave scattering amplitude in UPA is given by,
\begin{equation}
 f_3^{UPA}(p)=-\frac{m}{3\pi}T_3^{UPA}(p,p;E=\frac{3 p^2}{4 m}-\frac{\gamma^2}{m}). 
\label{f3}
\end{equation}

In Ref.~\cite{Shepard:2009}, this result is used to study the convergence of finite range expansions and compare with similar expansions using effective field theories.  Here we extend this to higher order in the separable expansion.

%%%%%%%%%%%%%%%%%%%%%%%%%%%%%%%%%%%%%%%%%%%%%%
\begin{figure}[ht]
\vspace{0.50in}
\includegraphics[width=6in,angle=0]{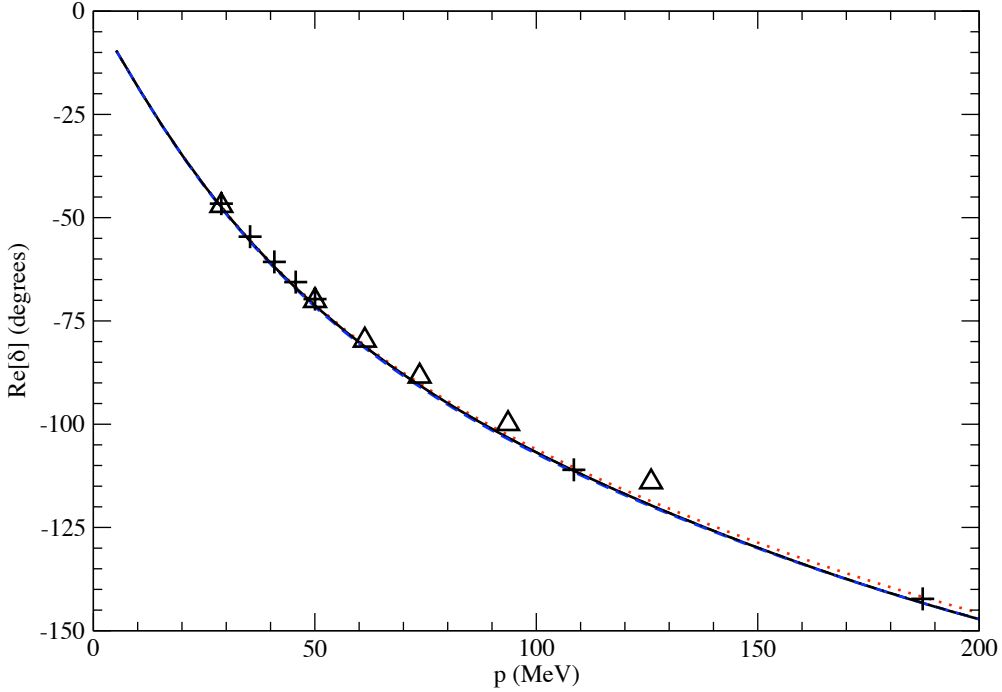}
\caption{(Color online).
The real part of $L=0$ $nd$ quartet phase shifts using three implementations of the rank-2 Harms separable expansion.  The solid  curve (black) is the rank-2 calculation for $V_{0}$ and the (indistinguishable) dashed curve (blue) is the rank-2 calculation for $V_{low}$. The dotted curve (red) is the rank-2 calculation using phenomenological (gaussian times polynomial) form factors fit to the 2-body phase shifts.  For this case the effect of reducing the cutoff from $\Lambda = 16$ fm$^{-1}$ to $\Lambda = 2$ fm$^{-1}$ has no significant effect. The triangles are the full Faddeev calculations of Huber, et al.~\cite{Huber:1993}  and the crosses are the calculations of Kievsky, et al.~\cite{Kievsky:2004-737} (and references therein).}
\label{Rank2_S_Re}
\end{figure}
%%%%%%%%%%%%%%%%%%%%%%%%%%%%%%%%%%%%%%%%%%%%%%

%%%%%%%%%%%%%%%%%%%%%%%%%%%%%%%%%%%%%%%%%%%%%%
\begin{figure}[ht]
\vspace{0.50in}
\includegraphics[width=6in,angle=0]{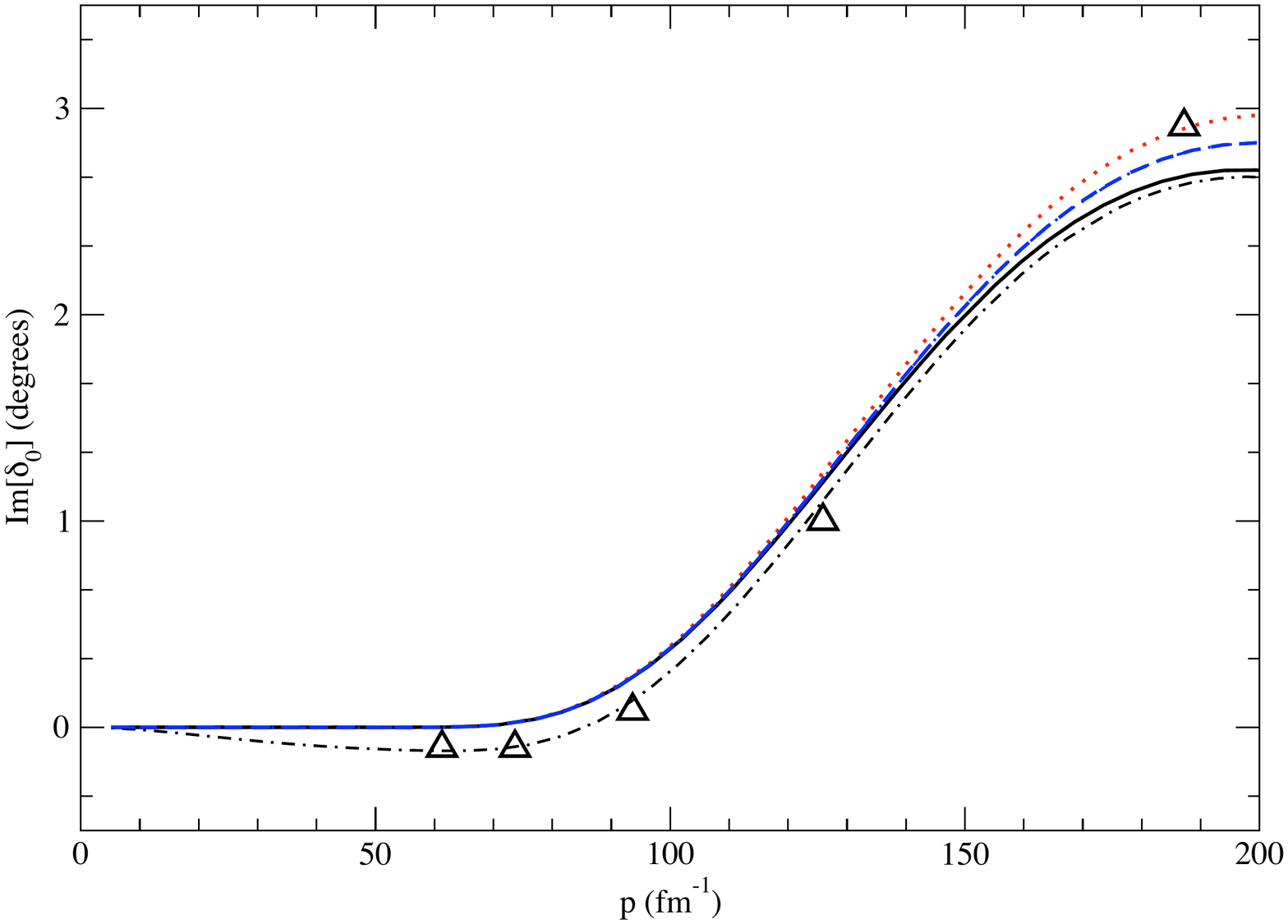}
\caption{(Color online).
The imaginary part of $L=0$ $nd$ quartet phase shifts using three implementations of the rank-2 Harms separable expansion.  The solid curve (black) is the rank-2 calculation using for $V_{0}$ and the dashed curve (blue) is the rank-2 calculation for $V_{low}$. The dotted curve (red) is the rank-2 calculation using phenomenological (gaussian times polynomial) form factors fit to the 2-body physics. Also shown is the effect of reducing the cutoff from $\Lambda = 16$ fm$^{-1}$ to $\Lambda = 2$ fm$^{-1}$.  This change has no effect on the $V_{low}$ calculation, but gives the dash-dot (black) curve for $V_0$ which violates unitarity below the breakup threshold at $p=53$ MeV/c.  The triangles are the full Faddeev calculations of Huber, et al.~\cite{Huber:1993} and the crosses are the calculations of Kievsky, et al.~\cite{Kievsky:2004-737} (and references therein).}
\label{Rank2_S_Im}
\end{figure}
%%%%%%%%%%%%%%%%%%%%%%%%%%%%%%%%%%%%%%%%%%%%%%
 
%%%%%%%%%%%%%%%%%%%%%%%%%%%%%%%%%%%%%%%%%%%%%%
\begin{figure}[ht]
\vspace{0.50in}
\includegraphics[width=6in,angle=0]{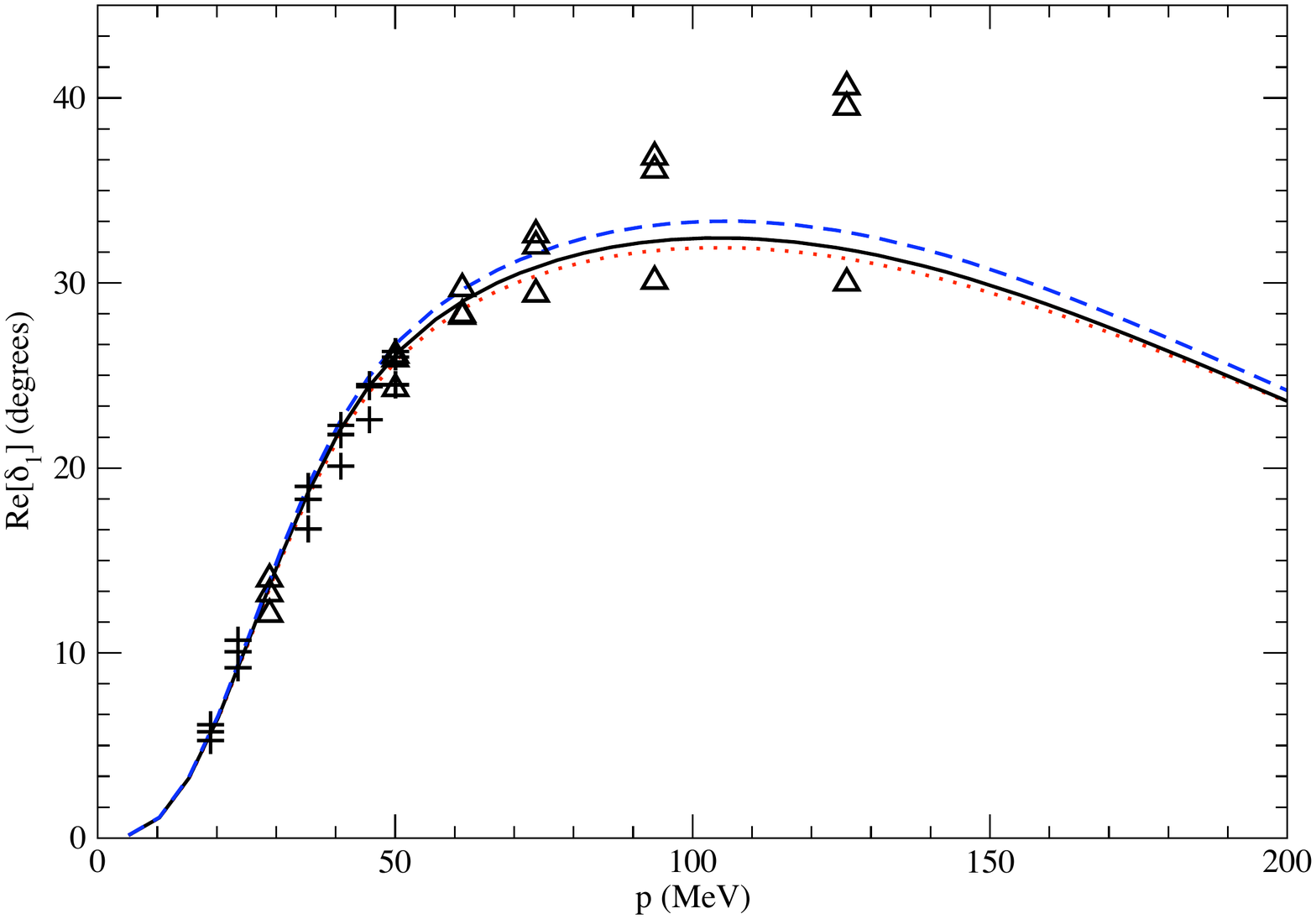}
\caption{(Color online).
The real part of $L=1$ $nd$ quartet phase shifts using three implementations of the rank-2 Harms separable expansion. The solid  curve (black) is the rank-2 calculation using for $V_{0}$ and the dashed curve (blue) is the rank-2 calculation for $V_{low}$. The dotted curve (red) is the rank-2 calculation using phenomenological (gaussian times polynomial) form factors fit to the 2-body phase shifts.  The triangles are the $L=1$  ($^4P_{1/2}$, $^4P_{3/2}$, $^4P_{5/2}$) full Faddeev calculations of Huber, et al.~\cite{Huber:1993}  and the crosses are the calculations of Kievsky, et al.~\cite{Kievsky:2004-737} (and references therein). }
\label{Rank2_P_Re}
\end{figure}
%%%%%%%%%%%%%%%%%%%%%%%%%%%%%%%%%%%%%%%%%%%%%%

%%%%%%%%%%%%%%%%%%%%%%%%%%%%%%%%%%%%%%%%%%%%%%
\begin{figure}[ht]
\vspace{0.50in}
\includegraphics[width=6in,angle=0]{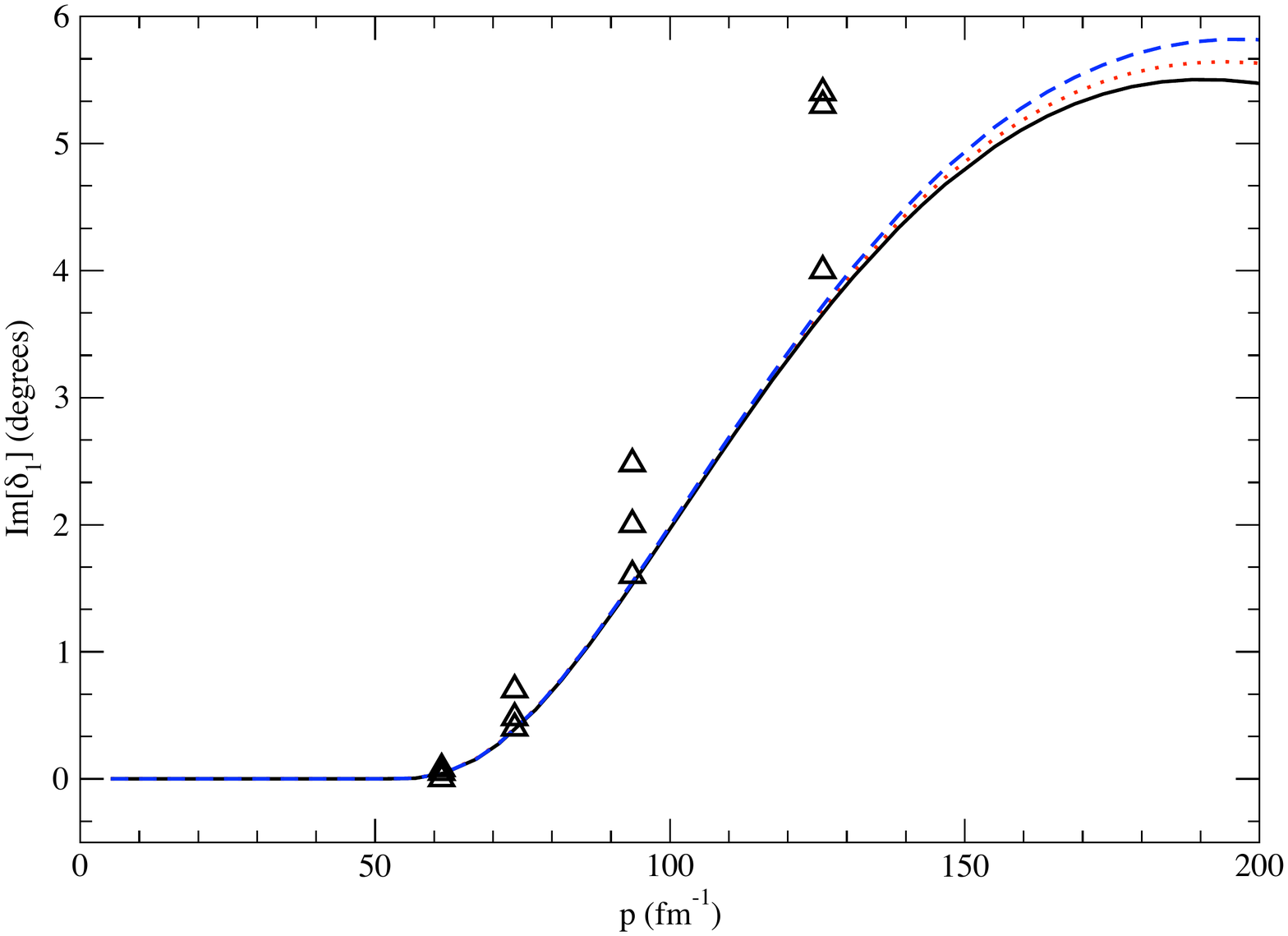}
\caption{(Color online).
         Same as for Fig.~\ref{Rank2_P_Re} but for the imaginary part of the phase shifts. The breakup threshold is at $p=53$ MeV/c.}
\label{Rank2_P_Im}
\end{figure}
%%%%%%%%%%%%%%%%%%%%%%%%%%%%%%%%%%%%%%%%%%%%%%

\section{Beyond the UPA}
\label{beyondUPA}

To move beyond the UPA we examine the 3-body scattering equations for separable potentials of arbitrary, but finite, rank as introduced in the 2-body context in Section~\ref{Sec:Harms}.  We closely follow the treatment of Watson and Nuttall with minor changes in notation. We restrict our attention to the scattering of one particle from the bound state (``dimer") of the other two.  We work in the center-of-mass system and assume that the particles have identical mass, $m$.  To begin, we recognize that the 2-body form factors now act in the 3-body Hilbert space; so a more elaborated notation is needed that allows us to identify the various 2-body partitions.  We adopt the ``odd-man-out" notation where the subscript $\bar\imath$ refers to the $\{jk\}$-pair where neither $j$ nor $k$ equals $i$, e.g. $\bar3$ refers to the $\{12\}$-pair.  Thus, the 3-body state with particles 1 and 2 in state $|\psi_n\rangle$ and particle 3 in state $|p_3\rangle$ will be represented by
\begin{equation}
|\psi_n(1,2)\rangle\otimes |{\bf p}_3\rangle\leftrightarrow |n_{\bar3},{\bf p}_3\rangle.
\label{psi3bdy}
\end{equation}
Let $E$ be the 3-body energy and $G^{(3)}_0(E)$ be the free 3-body propagator, then
\begin{equation}
\langle n'_{\bar \imath},{\bf p_i}' | G^{(3)}_0(E)|n_{\bar j},{\bf p_j}\rangle=(2\pi)^3 \delta^{(3)}({\bf p_i}'-{\bf p_j}) \langle n'_{\bar\imath} |G^{(2)}_0(E- p_i^2/2\mu_i)|n_{\bar j} \rangle,
\label{diagME1}
\end{equation}
where $G^{(2)}_0(z)$ is the free 2-body propagator and $\mu_i$ is the reduced mass of particle $i$ with the ${\bar \imath}$-dimer.  Under the assumption of equal masses, $\mu_i=2 m/3$ independent of $i$. If $B$ is the dimer binding energy, then the particles are on the energy shell when $E=3 p^2/4m-B$ and
\begin{equation}
\langle n'_{\bar \imath}| G^{(2)}_0(E-3 p^2/4m)|n_{\bar \imath} \rangle\rightarrow\langle n'_{\bar \imath} |G^{(2)}_0(-B)|n_{\bar \imath} \rangle=-\delta_{n'_{\bar \imath},n_{\bar \imath}}
\label{G2}
\end{equation}
as in Eq.~\ref{FFnorm1} of Section~\ref{Sec:Harms}.  The Faddeev Equation for the 3-body wave function for, e.g., $\bar\imath={\bar3}$ is (see {\it e.g.} Watson and Nuttall, Eq. 4.40)
\begin{eqnarray}
|\psi_{n_{\bar3},{\bf p}_3;E}\rangle&=&\left( \begin{matrix}{|\psi^{(1)}_{n_{\bar3},{\bf p}_3;E}\rangle}\\ {|\psi^{(2)}_{n_{\bar3},{\bf p}_3;E}\rangle}\\ {|\psi^{(3)}_{n_{\bar3},{\bf p}_3;E}\rangle}\end{matrix}\right) \nonumber\\ 
&=&G^{(3)}_0(E)\left( \begin{matrix}{0}\\ {0}\\ {|n_{\bar3},{\bf p}_3\rangle}\end{matrix}\right)
+G^{(3)}_0(E)\left( \begin{matrix}{0}&{T_{\bar1}(E)}&{T_{\bar1}(E)}\\ {T_{\bar2}(E)}&{0}&{T_{\bar2}(E)}\\ {T_{\bar3}(E)}&{T_{\bar3}(E)}&{0}\end{matrix}\right)\left( \begin{matrix}{|\psi^{(1)}_{n_{\bar3},{\bf p}_3; E}\rangle}\\ {|\psi^{(2)}_{n_{\bar3},{\bf p}_3; E}\rangle}\\ {|\psi^{(3)}_{n_{\bar3},{\bf p}_3; E}\rangle}\end{matrix}\right)\nonumber\\
\label{Faddeev0}
\end{eqnarray}
where in $|\psi^{(i)}_{n_{\bar3},{\bf p}_3; E}\rangle$, the ${\bar\imath}$-pair interact last so the $i$-th particle is free asymptotically with momentum ${\bf p}_3$ and where $T_{\bar\imath}(E)$ are 2-body T-matrices for the ${\bar\imath}$-pair acting in the 3-body space which for a rank-N separable potential are given by,
\begin{equation}
T_{\bar\imath}(E)=-\sum_{n_{\bar\imath},m_{\bar\imath}}^N\int\frac{d^3p''_i}{(2\pi)^3}\ |n_{\bar\imath},{\bf p}_i''\rangle\ \Delta_{n_{\bar\imath} m_{\bar\imath}}(E-3  {p''_i}^2/4m)\ \langle m_{\bar\imath},{\bf p}_i''|
\label{T2}
\end{equation}
where $\Delta_{nm}(z)$ appears in Eq.~\ref{LSeq2}. Hence,
\begin{eqnarray}
|\psi^{(1)}_{n_{\bar3},{\bf p}_3; E}\rangle&=&-G^{(3)}_0(E)\sum_{l_{\bar1},m_{\bar1}}^N\int\frac{d^3p''_1}{(2\pi)^3}\ |l_{\bar1},{\bf p}_1''\rangle\ \Delta_{l_{\bar1}m_{\bar1}}(E-3{{ p}_1''}^2/4m) \nonumber\\
&&\qquad\qquad \times\biggl[\langle m_{\bar1},{\bf p}_1''|\psi^{(2)}_{n_{\bar3},{\bf p}_3; E}\rangle +\langle m_{\bar1},{\bf p}_1''|\psi^{(3)}_{n_{\bar3},{\bf p}_3; E}\rangle \biggr], 
\label{Faddeev1a}
\end{eqnarray}
\begin{eqnarray}
|\psi^{(2)}_{n_{\bar3},{\bf p}_3; E}\rangle&=&-G^{(3)}_0(E)\sum_{l_{\bar2},m_{\bar2}}^N\int\frac{d^3p''_2}{(2\pi)^3}\ |l_{\bar2},{\bf p}_2''\rangle\ \Delta_{l_{\bar2}m_{\bar2}}(E-3{{ p}_2''}^2/4m) \nonumber\\
&&\qquad\qquad \times\biggl[\langle m_{\bar2},{\bf p}_2''|\psi^{(1)}_{n_{\bar3},{\bf p}_3; E}\rangle +\langle m_{\bar2},{\bf p}_2''|\psi^{(3)}_{n_{\bar3},{\bf p}_3; E}\rangle \biggr], 
\label{Faddeev1b}
\end{eqnarray}
and
\begin{eqnarray}
|\psi^{(3)}_{n_{\bar3},{\bf p}_3; E}\rangle&=&G^{(3)}_0(E)\biggl\{|n_{\bar3},{\bf p}_3\rangle-\sum_{l_{\bar3},m_{\bar3}}^N\int\frac{d^3p''_3}{(2\pi)^3}\ |l_{\bar3},{\bf p}_3''\rangle\ \Delta_{l_{\bar3}m_{\bar3}}(E-3{{ p}_3''}^2/4m) \nonumber\\
&&\qquad\qquad \times\biggl[\langle m_{\bar3},{\bf p}_3''|\psi^{(1)}_{n_{\bar3},{\bf p}_3;E}\rangle +\langle m_{\bar3},{\bf p}_3''|\psi^{(2)}_{n_{\bar3},{\bf p}_3;E}\rangle \biggr]\biggr\}. 
\label{Faddeev1c}
\end{eqnarray}
Now define the 3-body amplitudes,
\begin{equation}
X_{n'_{\bar\imath} n_{\bar j}}({\bf p'}_\imath,{\bf p}_j; E)=\sum_{k\neq j}\langle n'_{\bar\imath},{\bf p}_\imath'|\psi^{(k)}_{n_{\bar j},{\bf p}_j; E}\rangle,
\end{equation}% X33-def
to find explicitly,
\begin{eqnarray}
X_{n'_{\bar3} n_{\bar3}}({{\bf p}_3',{\bf p}_3; E})=&&-\sum_{l_{\bar1},m_{\bar1}}\int\frac{d^3p_1''}{(2\pi)^3}\ \langle n'_{\bar3},{\bf p}_3'|G^{(3)}_0(E)| l_{\bar1},{\bf p}_1''\rangle\ \Delta_{ l_{\bar1} m_{\bar1}}(E-3 {p_1''}^2/4m)\ X_ { m_{\bar1}n_{\bar3}}({{\bf p}_1'',{\bf p}_3; E})\nonumber\\ 
&&-\sum_{ l_{\bar2}, m_{\bar2}}\int\frac{d^3p_2''}{(2\pi)^3}\ \langle n'_{\bar3},{\bf p}_3'|G^{(3)}_0(E)| l_{\bar2},{\bf p}_2''\rangle\ \Delta_{ l_{\bar2} m_{\bar2}}(E-3 {p_2''}^2/4m)\ X_{ m_{\bar2}n_{\bar3}}({{\bf p}_2'',{\bf p}_3; E}),\nonumber \\
\label{X33}
\end{eqnarray} % X33
\begin{eqnarray}% X13
X_{n'_{\bar1}n_{\bar3}}({{\bf p}_1' ,{\bf p}_3; E})&&=\langle n'_{\bar1},{\bf p}_1'|G^{(3)}_0(E)|n_{\bar3},{\bf p}_3\rangle \nonumber\\ 
&&-\sum_{l_{\bar2},m_{\bar2}}\int\frac{d^3p_2''}{(2\pi)^3}\ \langle n'_{\bar1},{\bf p}_1'|G^{(3)}_0(E)| l_{\bar2},{\bf p}_2''\rangle\ \Delta_{ l_{\bar2} m_{\bar2}}(E-3 {p_2''}^2/4m)\ X_{ m_{\bar2}n_{\bar3}} ({{\bf p}_2'',{\bf p}_3; E}) \nonumber\\
&&-\sum_{ l_{\bar3}, m_{\bar3}}\int\frac{d^3p_3''}{(2\pi)^3}\ \langle n'_{\bar1},{\bf p}_1'|G^{(3)}_0(E)| l_{\bar3},{\bf p}_3''\rangle\ \Delta_{ l_{\bar3} m_{\bar3}}(E-3 {p_3''}^2/4m)\ X_{ m_{\bar3}n_{\bar3}}({{\bf p}_3'',{\bf p}_3; E}), \nonumber\\
\label{X13}% X13
\end{eqnarray}
and
\begin{eqnarray}% X23
X_{n'_{\bar2}n_{\bar3}}({{\bf p}_2',{\bf p}_3; E})&&=\langle n'_{\bar2},{\bf p}_2'|G^{(3)}_0(E)|n_{\bar3},{\bf p}_3\rangle \nonumber\\ 
&&-\sum_{l_{\bar1},m_{\bar1}}\int\frac{d^3p_1''}{(2\pi)^3}\ \langle n'_{\bar2},{\bf p}_2'|G^{(3)}_0(E)| l_{\bar1},{\bf p}_1''\rangle\ \Delta_{ l_{\bar1} m_{\bar1}}(E-3 {p_1''}^2/4m)\ X_{ m_{\bar1}n_{\bar3}}({{\bf p}_1'',{\bf p}_3; E}) \nonumber\\ 
&&-\sum_{ l_{\bar3}, m_{\bar3}}\int\frac{d^3p_3''}{(2\pi)^3}\ \langle n'_{\bar2},{\bf p}_2'|G^{(3)}_0(E)| l_{\bar3},{\bf p}_3''\rangle\ \Delta_{ l_{\bar3} m_{\bar3}}(E-3 {p_3''}^2/4m)\ X_{ m_{\bar3}n_{\bar3}}({{\bf p}_3'',{\bf p}_3; E}).\nonumber\\
\label{X23}% X23
\end{eqnarray}
So far, the three particles have been treated as distinguishable. Now we assume they are identical spinless bosons in which case we may, without loss of generality, let
\begin{equation}
X_{n'_{\bar\imath}n_{\bar j}}({{\bf p}_\imath',{\bf p}_j; E})\rightarrow \delta_{\imath j}X_{n'n}^D({\bf p}',{\bf p}; E)
\label{X_D}
\end{equation} %X_D
and
\begin{equation}
X_{n'_{\bar\imath} n_{\bar j}}({{\bf p}_\imath',{\bf p}_j; E}),\ X_{n'_{\bar\imath},n_{\bar j}}({{\bf p}_\imath' ,{\bf p}_j; E})\rightarrow (1-\delta_{ij})X_{n'n}^X({\bf p}', {\bf p}; E)
\label{X_X}
\end{equation}  %X_X
where $X^D$ and $X^X$ are ``direct" and ``exchange" amplitudes, respectively.  We then have
\begin{equation}
X_{n'n}^D({\bf p}', {\bf p}; E)=
-2\sum_{l,m}^N\int\frac{d^3p''}{(2\pi)^3}\ \langle n',{\bf p}'|G^{(3)}_0(E)|l,{\bf p}''\rangle_X\ \Delta_{lm}(E-3 {p''}^2/4m)\ X_{mn}^X({\bf p}'', {\bf p}; E),
\label{Faddeev_D}
\end{equation} %X_Deqn
and
\begin{eqnarray}
X_{n'n}^X({\bf p}', {\bf p}; E)&=&  \langle n',{\bf p}'|G^{(3)}_0(E)|n,{\bf p}\rangle_X \nonumber\\ 
&&-\sum_{l,m}^N\int\frac{d^3p''}{(2\pi)^3}\ \langle n',{\bf p}'|G^{(3)}_0(E)|l,{\bf p}''\rangle_X\ \Delta_{lm}(E-3 {p''}^2/4m)\nonumber\\ 
&&\qquad\qquad\qquad\qquad \times\biggl[ X_{mn}^D({\bf p}'', {\bf p}; E) + X_{mn}^X({\bf p}'', {\bf p}; E) \biggr] 
\label{Faddeev_X}
\end{eqnarray} %X_Xeqn
where the ``$X$" subscript in $\langle n',{\bf p}'|G^{(3)}_0(E)| n,{\bf p}\rangle_X$ means that the bra and ket must correspond to different arrangements of the three bosons as required by Eq.~\ref{X33}-\ref{X23}.
Then define $X=X^D+2 X^X$, whence the two previous equations can be combined to yield
\begin{eqnarray}
&&X_{n'n}({\bf p}', {\bf p}; E)= 2\langle n',{\bf p}'|G^{(3)}_0(E)|n,{\bf p}\rangle_X \\ \nonumber
&&-2\sum_{l,m}\int\frac{d^3p''}{(2\pi)^3}\ \langle n',{\bf p}'|G^{(3)}_0(E)|l,{\bf p}''\rangle_X\ \Delta_{lm}(E-3 {p''}^2/4m)\ X_{mn}({\bf p}'', {\bf p}, E).
\label{X_DX}
\end{eqnarray} %X_DX
We now write
\begin{equation}
\langle n',{\bf p}'|G^{(3)}_0(E)|n,{\bf p}\rangle_X=\frac{\langle\psi_{n'}|{\bf p}'+\frac{\bf p}{2}\rangle
\langle{\bf p}+\frac{\bf p'}{2}|\psi_n\rangle}{E-\frac{1}{m}(p^2+{p'}^2+{\bf p}'\cdot{\bf p})},
\label{FullBorn1} %Born
\end{equation}
and then define partial wave components,
\begin{eqnarray}
{\cal Z}_{n'n}^{L}(p', p; E)&=&\frac{1}{2}\int_{-1}^{+1} dx\ P_L(x)\ \langle n',{\bf p}'|G^{(3)}_0(E)|n,{\bf p}\rangle_X \nonumber \\
&=&\frac{1}{2}\int_{-1}^{+1} dx\ P_L(x)\ \frac{\langle\psi_{n'}|\sqrt{{p'}^2/4+p^2+p' p x}\rangle\langle\sqrt{{p'}^2+p^2/4+p' p x}|\psi_n\rangle}{E-\frac{1}{m}({p'}^2+p^2+p' p x)}\nonumber \\
\label{Z_L} % Z_L
\end{eqnarray}
where $P_L(x)$ is a Legendre polynomial.  There are then, for a rank-$N$ separable potential, $N^2$ coupled Faddeev equations for bosonic scattering of the form,
\begin{eqnarray}
&&X_{n'n}^L(p', p; E)= 2 {\cal Z}_{n'n}^L(p', p; E) \nonumber\\
&&-\frac{2}{2\pi^2}\sum_{l,m}^N\int_0^\infty {p''}^2 dp''\  {\cal Z}_{n'l}^L(p', p''; E)\ \Delta_{lm}(E-3 {p''}^2/4m)\ X_{mn}^L(p'', p; E).
\label{X_L}
\end{eqnarray}
The corresponding equations for $nd$ quartet scattering are obtained by the simple expedient of multiplying the ${\cal Z}$'s by the isospin-spin structure factor $\Lambda_{IS}=-1/2$ (see Eqns.~\ref{Z} and~\ref{SI}). 

We now consider the special case of a rank-2 separable potential, and we observe that the $2^2$ equations decouple into two pairs of coupled equations. We will focus on elastic scattering, so, assuming that $n=1$ corresponds to the deuteron ground state, we focus on the equations for $X_{11}$ and $X_{21}$:
\begin{eqnarray}
X_{11}^L(p' , p;E)=&& 2 \Lambda_{IS} {\cal Z}_{11}^L(p',p;E)\nonumber \\ 
&&-\frac{2 \Lambda_{IS}}{2\pi^2}\int_0^\infty {p''}^2 d{p''}\  \biggl[{\cal Z}_{11}^L(p',{p''};E)\ \Delta_{11}(E-3 {p''}^2/4m)  \nonumber \\
&&\qquad\qquad\qquad\qquad +{\cal Z}_{12}^L(p',{p''};E)\ \Delta_{21}(E-3 {p''}^2/4m)\biggr] X_{11}^L({p''},p;E) \nonumber \\ 
&&-\frac{2 \Lambda_{IS}}{2\pi^2}\int_0^\infty {p''}^2 d{p''}\  \biggl[{\cal Z}_{11}^L(p',{p''};E)\ \Delta_{12}(E-3 {p''}^2/4m) \nonumber\\
&&\qquad\qquad\qquad\qquad+ {\cal Z}_{12}^L(p',{p''};E)\ \Delta_{22}(E-3 {p''}^2/4m)\biggr]  X_{21}^L({p''},p;E),\nonumber\\
\label{X_11}
\end{eqnarray}
and
\begin{eqnarray}
X_{21}^L(p' , p;E)=&& 2 \Lambda_{IS} {\cal Z}_{21}^L(p',p;E)\nonumber\\ 
&&-\frac{2 \Lambda_{IS}}{2\pi^2}\int_0^\infty {p''}^2 d{p''}\  \biggl[{\cal Z}_{21}^L(p',{p''};E)\ \Delta_{11}(E-3 {p''}^2/4m) \nonumber \\
&&\qquad\qquad\qquad\qquad+ {\cal Z}_{22}^L(p',{p''};E)\ \Delta_{21}(E-3 {p''}^2/4m)\biggr] X_{11}^L({p''},p;E)\nonumber \\ 
&&-\frac{2 \Lambda_{IS}}{2\pi^2}\int_0^\infty {p''}^2 d{p''}\  \biggl[{\cal Z}_{21}^L(p',{p''};E)\ \Delta_{12}(E-3 {p''}^2/4m) \nonumber \\ 
&&\qquad\qquad\qquad\qquad+ {\cal Z}_{22}^L(p',{p''};E)\ \Delta_{22}(E-3 {p''}^2/4m)\biggr]  X_{21}^L({p''},p;E).\nonumber\\
\label{X_12}
\end{eqnarray}
The $nd$ quartet elastic scattering amplitude for the L-th partial wave is
\begin{equation}
f^L(p,p;E=\frac{3p^2}{4 m}-B)=-\frac{m}{3\pi}\ Z_1\ X_{11}^L(p,p;E=\frac{3p^2}{4 m}-B),
\label{f_L}
\end{equation}
where $Z_1=Z_B$ is the bound state normalization given by Eq.~\ref{Zbnd}.

%%%%%%%%%%%%%%%%%%%%%%%%%%%%%%%%%%%%%%%%%%%%%%
\section{Results}
\label{Sec:results}

We begin with calculations of the  $^4$S$_{\frac{3}{2}}$ phase shifts using three implementations of the rank-1, unitary pole approximation (UPA).  Shown in Figs.~\ref{UPA_S_Re} and \ref{UPA_S_Im} are calculations of the real and imaginary $^4$S$_{\frac{3}{2}}$ phase shifts for our three implementations of the UPA compared to the full Faddeev calculations of Huber, et al.~\cite{Huber:1993} and Kievsky, et al.~\cite{Kievsky:2004-737} and references therein.  The three implementations are (1) the phenomenological dipole form factor treated in Section~\ref{Sec:UPA}, (2) the rank-1 Harms form factor for the bare Malfleit-Tjon triplet potential ($V_0$), and (3) the rank-1 form factor for $V_{low}$ which is obtained through a similarity renormalization group evolution of $V_0$ from 16 fm$^{-1}$ to 2 fm$^{-1}$ as described in Section~\ref{Sec:SRG}.  All three implementations accurately reproduce the real part of the s-wave quartet phase shift of the full Faddeev calculation, and each method gives a respectable reproduction of the imaginary S-wave phase shifts when the cutoff is large, $\Lambda_0 = 16$ fm$^{-1}$ .  Also shown is the effect of reducing the cutoff momentum in the 3-body calculation from $\Lambda_0 = 16$ fm$^{-1}$ to $\Lambda_0 = 2$ fm$^{-1}$.  The real part of the phase shift is relatively insensitive to this change.  Reducing the cut-off momentum causes the imaginary part of the phase shift to be reduced somewhat at the larger momenta, and, in the case of $V_0$ introduces a spurious negative imaginary phase shift at low momenta. No such spurious effect is seen using $V_{low}$ however.

Shown in Figs.~\ref{Rank2_S_Re} and \ref{Rank2_S_Im} are calculations of the real and imaginary $^4$S$_{\frac{3}{2}}$ phase shifts for three implementations of rank-2 separable expansion compared to the full Faddeev calculations of Huber, et al.~\cite{Huber:1993} and Kievsky, et al.~\cite{Kievsky:2004-737} and references therein.  The three implementations are  (1) a phenomenological form factor (polynomial times gaussian) fit to the NN phase shifts, (2) the rank-2 Harms expansion using the first two form factors for the bare Malfleit-Tjon triplet potential ($V_0$) and (3) the rank-2 Harms expansion using the first two form factors for $V_{low}$ which is obtained through a similarity renormalization group evolution of $V_0$ from 16 fm$^{-1}$ to 2 fm$^{-1}$ as described in Section~\ref{Sec:SRG}.  The form factors are shown in Fig.~\ref{Fig2}.  Each method gives a respectable reproduction of the full Faddeev calculations for both the real and imaginary S-wave phase shifts.  Also shown is the effect of reducing the cut-off momentum in the 3-body calculation from $\Lambda_0 = 16$ fm$^{-1}$ to $\Lambda_0 = 2$ fm$^{-1}$.   Similar to the rank-1 case, we see that reducing the cut-off momentum does not significantly affect the real part of the phase shift and causes the imaginary part of the phase shift to be reduced at the larger momenta but less so than in the rank-1 case, and, as in the rank-1 case, introduces a small spurious negative imaginary phase shift at low momenta for the $V_{0}$ implementation.  We conclude that the high momentum contributions are necessary to retain unitarity below threshold for the $V_0$ case, but these contributions are included in $V_{low}$ such that unitarity is respected even when the cut-off is reduced. 

We have also calculated higher partial waves and find similar levels of agreement and systematics. Shown in Figs.~\ref{Rank2_P_Re} and \ref{Rank2_P_Im} are calculations of the real and imaginary $^4$P$_{j}$ phase shifts for three implementations of rank-2 separable expansion compared to the full Faddeev calculations of Huber, et al.~\cite{Huber:1993} and Kievsky, et al.~\cite{Kievsky:2004-737} and references therein. Since our simple model does not have spin-orbit coupling, our phase shifts are independent of the $j$-value.  Nevertheless, the calculations of the real part of the phase shift lies within the range of results from the full Faddeev calculations including spin-orbit forces while the results for the small imaginary part of the phase shift fall just below those of the full Faddeev calculations at the higher momenta.

%%%%%%%%%%%%%%%%%%%%%%%%%%%%%%%%%%%%%%%%%%%%%%
\begin{table}
\begin{tabular}{|l|c|c|c|c|}
 \hline
~~~~~~~~~~~Model &~~~~~~$\sigma_{tot}$ (mb)~~& ~~~~~~$\sigma_{el}$ (mb)~~ &~~~~~ $\sigma_{BR}$ (mb)~~ &~ $\frac{d\sigma}{d\Omega}(\theta=180^o)$ (mb)\\
 \hline
UPA (Form Factor)& 752.3 (751.6) & 712.9 (712.2) & 39.4 (39.4) & 177.6  (177.4)\\
UPA ($V_0$) & 776.4 (777.7)& 736.9 (739.6) & 39.5 (38.1) & 179.8 (180.5) \\
UPA ($V_{low}$)& 789.6 (789.5)& 749.6 (749.6) & 40.0 (40.0) &182.0 (182.0) \\
Rank-2 (Gaussian x polynomial)& 770.9 (770.9)& 730.9 (731.3) & 40.0 (39.6)& 179.2 (179.4)\\
Rank-2 Harms ($V_0$) & 779.0 (780.3) & 739.3 (742.1)& 39.6 (38.2)& 180.9 (181.6) \\
Rank-2 Harms ($V_{low}$) & 794.4 (794.3)& 754.3 (754.2)& 40.1 (40.1)& 184.2 (184.2) \\
Koike \&Yaniguchi \cite{Koike:1986}& 812.3 & 770.2 & 42.1 & 185.0 \\
\hline
\end{tabular}
\caption{Calculations of center of mass total, elastic, break-up, and back-angle cross sections for quartet $nd$ scattering at $E_{lab}=10$ MeV, summing partial waves to $L_{max}=8$ using a cutoff of $\Lambda_0 = 16$ fm$^{-1}$. (The results from reducing the cutoff to $\Lambda_0 = 2$ fm$^{-1}$  are given in parentheses.)}
\label{Table2}
\end{table}
%%%%%%%%%%%%%%%%%%%%%%%%%%%%%%%%%%%%%%%%%%%%%%

 In Table~\ref{Table2} we present calculations of the total, elastic, break-up, and back-angle cross sections for quartet nd scattering at E$_{lab}$ = 10 MeV summing partial wave contributions to $L_{max} = 8$ compared with the results of Koike and Yaniguchi~\cite{Koike:1986}.  The back-angle $nd$ cross section is dominated by the quartet term.  As already observed long ago, for these quartet-dominated scattering measurements, the UPA results are quite respectable, especially considering their simplicity.  The quality and similarity of the three rank-1 calculations is somewhat surprising considering how different are the various rank-1 form factors, shown in Fig.~\ref{Fig2}.  Apparently, including the correct 2-body phenomenology through the leading form factor accounts for most of the physics involved in quartet scattering.  Still, it is satisfying that including the rank-2 term systematically improves the results for all cases considered and the rank-2 Harms expansion of the SRG-evolved potential, $V_{low}$, yields the best results compared with full Faddeev calculations.  We also note that the cross sections calculated using the rank-2 Harms expansion for $V_{low}$ are largely insensitive to the cutoff, giving essentially identical results down to $\Lambda_0 =$ 2 fm$^{-1}$.

\section{Conclusions}
\label{Sec:conclusions}

In this work we reviewed Harms' method of separable expansions and, for the first time, applied it to a potential evolved to low momentum using the similarity renormalization group.  This work differs significantly from that of Kamada, et al. (who use Legendre polynomials for the separable form factors of an RG-evolved potential) in that our Harms form factors, derived from the 2-body potential, incorporate the 2-body physics, and in particular the bound state physics, in the leading term. The goal of this work was to develop a computationally simple yet accurate approach to 3-body calculations.  To demonstrate the utility and feasibility of this approach while avoiding the complications of tensor coupling and Efimov physics, we applied the Harm's separable expansion to 3-body nd quartet scattering using both the bare Malfliet-Tjon-III potential, $V_0$, and its SRG-evolved potential, $V_{low}$.  We find that excellent agreement with full Faddeev calculations can be achieved at just rank-2 and, not surprisingly, that the results using $V_{low}$ are stable with respect to the momentum cutoff avoiding spurious non-physical negative imaginary phase shifts that arise for $V_0$ when the cutoff is lowered.

The $nd$ quartet case is an important proof-of-principle example but has has long been known to be well described by the unitary pole approximation, so the quality of our rank-1 results and improvements from higher rank terms may not seem surprising.  The real test of our approach will come in future applications to the nd doublet case which is theoretically more challenging due to complications of tensor coupling and Efimov physics.

\smallskip
\acknowledgments
One of the authors (JAM) thanks the Physics Department of the University of Colorado, Boulder for hosting him during his sabbatical where work on this project was initiated.  This work was supported in part by DOE grant DE-FG05-92ER40750.

\vfill\eject
\bibliography{References.bib}

\end{document}